\documentclass[longauth]{aa}
\usepackage{txfonts}
\usepackage{graphicx}	
\usepackage{amsmath}	
\usepackage{mathrsfs}
\usepackage[breaklinks, colorlinks, citecolor=blue, urlcolor = blue]{hyperref}
\usepackage{pdflscape}
\usepackage{float}
\usepackage{here}
\usepackage[export]{adjustbox}
\usepackage{array}
\usepackage{natbib}
\bibpunct{(}{)}{;}{a}{}{,} 
\newcommand{\xmm}{\textit{XMM-Newton }}
\newcommand{\mseed}{$M_{\rm s, Edd}$}
\newcommand{\logmseed}{$\log(M_{\rm s, Edd})$ }

\defcitealias{Zappacosta2023}{Paper~I}
\begin{document} 

   \title{HYPERION. Shedding light on the first luminous quasars: A correlation between UV disc winds and X-ray continuum.}
   \titlerunning{HYPERION. Correlation between UV disc winds and X-ray continuum}
   \authorrunning{Tortosa et al.}


   \author{A. Tortosa\thanks{\email{alessia.tortosa@inaf.it}}\inst{1}
          \and L. Zappacosta \inst{1}
          \and E. Piconcelli \inst{1}
          \and M. Bischetti \inst{2,3}
          \and C. Done \inst{4}
          \and G. Miniutti \inst{5}
          \and I. Saccheo \inst{6,1}
          \and G. Vietri \inst{7}
          \and A. Bongiorno \inst{1}
          \and M. Brusa \inst{8,9}
          \and S. Carniani \inst{10}
          \and I. V. Chilingarian \inst{14} 
          \and F. Civano \inst{11}
          \and S. Cristiani \inst{2,12,13}
          \and V. D'Odorico \inst{2,10,12}
          \and M. Elvis \inst{14}
          \and X. Fan \inst{15}
          \and C. Feruglio \inst{2,12}
          \and F. Fiore \inst{2,12}
          \and S. Gallerani \inst{10}
          \and E. Giallongo \inst{1}
          \and R. Gilli \inst{9}
          \and A. Grazian \inst{16}
          \and M. Guainazzi \inst{17}
          \and F. Haardt \inst{18,19,20}
          \and A. Luminari \inst{21,1}
          \and R. Maiolino \inst{22,23,24}
          \and N. Menci \inst{1}
          \and F. Nicastro \inst{1}
          \and P. O. Petrucci \inst{25}
          \and S. Puccetti \inst{26}
          \and F. Salvestrini \inst{2}
          \and R. Schneider \inst{1,27,28,29}
          \and V. Testa \inst{1}
          \and F. Tombesi \inst{30,31,1,32,11}
          \and R. Tripodi \inst{33,2,3,12}
          \and R. Valiante \inst{1,28}
          \and L. Vallini \inst{9}
          \and E. Vanzella \inst{9}
          \and A. Vasylenko \inst{34}
          \and C. Vignali \inst{8,9}
          \and F. Vito \inst{9}
          \and M. Volonteri \inst{35}
          \and F. La Franca \inst{6,1}
          }
   \institute{INAF - Osservatorio astronomico di Roma, Via Frascati 33, I-00040 Monte Porzio Catone, Italy.
   \and INAF - Osservatorio Astronomico di Trieste, Via G. Tiepolo 11, I-34143 Trieste, Italy.
   \and Dipartimento di Fisica, Sezione di Astronomia, Università di Trieste, via Tiepolo 11, I-34143 Trieste, Italy.
   \and Centre for Extragalactic Astronomy, Department of Physics, Durham University, South Road, Durham DH1 3LE, UK.
   \and Centro de Astrobiología (CAB), CSIC-INTA, Camino Bajo del Castillo s/n, ESAC campus, 28692 Villanueva de la Cañada, Spain.
   \and Dipartimento di Matematica e Fisica, Università Roma Tre, Via della Vasca Navale 84, 00146 Roma, Italy.
   \and INAF - Istituto di Astrofisica Spaziale e Fisica Cosmica Milano, Via A. Corti 12, 20133 Milano, Italy.
   \and Dipartimento di Fisica e Astronomia ‘Augusto Righi’, Università degli Studi di Bologna, via P. Gobetti, 93/2, 40129 Bologna, Italy.
   \and INAF - Osservatorio di Astrofisica e Scienza dello Spazio di Bologna, via Piero Gobetti, 93/3, I-40129 Bologna, Italy.
   \and Scuola Normale Superiore, Piazza dei Cavalieri 7, I-56126 Pisa, Italy.
   \and NASA Goddard Space Flight Center, Greenbelt, MD 20771, USA.
   \and IFPU - Institute for Fundamental Physics of the Universe, via Beirut 2, I-34151 Trieste, Italy.
   \and INFN - National Institute for Nuclear Physics, via Valerio 2, I-34127 Trieste, Italy.
   \and Center for Astrophysics - Harvard \& Smithsonian, Cambridge, MA 02138, USA.
   \and Steward Observatory, University of Arizona, Tucson, Arizona, USA.
   \and INAF - Osservatorio Astronomico di Padova, Vicolo dell'Osservatorio 5, I-35122, Padova, Italy.
   \and European Space Agency, ESTEC, Keplerlaan 1, 2201 AZ Noordwijk, The Netherlands
   \and DiSAT, Università degli Studi dell’Insubria, Via Valleggio 11, I-22100 Como, Italy.
   \and INFN - Sezione di Milano-Bicocca, Piazza della Scienza 3, I-20126 Milano, Italy.
   \and INAF - Osservatorio Astronomico di Brera, Via E. Bianchi 46, I-23807 Merate, Italy.
   \and INAF - Istituto di Astrofisica e Planetologia Spaziali, Via del Fosso del Caveliere 100, I-00133 Roma, Italy.
   \and Cavendish Laboratory, University of Cambridge, 19 J. J. Thomson Ave., Cambridge CB3 0HE, UK.
   \and Kavli Institute for Cosmology, University of Cambridge, Madingley Road, Cambridge CB3 0HA, UK.
   \and Department of Physics \& Astronomy, University College London, Gower Street, London WC1E 6BT, UK.
   \and Univ. Grenoble Alpes, CNRS, IPAG, F-38000 Grenoble, France.
   \and ASI - Agenzia Spaziale Italiana, Via del Politecnico snc, I-00133 Roma, Italy.
   \and Dipartimento di Fisica, Università di Roma La Sapienza, Piazzale Aldo Moro 2, I-00185 Roma, Italy.
   \and INFN - Sezione Roma1, Dipartimento di Fisica, Università di Roma La Sapienza, Piazzale Aldo Moro 2, I-00185 Roma, Italy.
   \and Sapienza School for Advanced Studies, Viale Regina Elena 291, I- 00161 Roma, Italy. 
   \and Physics Department, Tor Vergata University of Rome, Via della Ricerca Scientifica 1, 00133 Rome, Italy.
   \and INFN - Rome Tor Vergata, Via della Ricerca Scientifica 1, 00133 Rome, Italy.
   \and Department of Astronomy, University of Maryland, College Park, MD 20742, USA.
   \and University of Ljubljana, Department of Mathematics and Physics, Jadranska ulica 19, SI-1000 Ljubljana, Slovenia.
   \and Main Astronomical Observatory, National Academy of Sciences of Ukraine, 27 Akademika Zabolotnoho St., Kyiv 03143, Ukraine.
   \and Institut d’Astrophysique de Paris, Sorbonne Université, CNRS, UMR 7095, 98 bis bd Arago, 75014 Paris, France.
   }
   \date{Received XX, 2024; accepted YY, 2024}
\abstract
{One of the main open questions in the field of luminous ($L_{\rm bol}>10^{47}$\,erg\,s$^{-1}$) quasars (QSOs) at $z \gtrsim 6$ is the rapid formation ($< 1$\,Gyr) of their supermassive black holes (SMBHs). For this work we analysed the relation between the X-ray properties and other properties describing the physics and growth of both the accretion disc and the SMBH in QSOs at the Epoch of Reionization (EoR). The sample consists of 21 $z>6$ QSOs, which includes 16 sources from the rapidly grown QSOs from the HYPERION sample and five other luminous QSOs with available high-quality archival X-ray data. We discovered a strong and statistically significant ($>3\sigma$) relation between the X-ray continuum photon index ($\Gamma$) and the C\,\textsc{iv} disc wind velocity ($v_{\rm C\,\textsc{iv}}$) in $z>6$ luminous QSOs, whereby the higher the $v_{\rm C\,\textsc{iv}}$, the steeper the $\Gamma$. This relation suggests a link between the disc--corona configuration and the kinematics of disc winds. Furthermore, we find evidence at $>2-3\sigma$ level that $\Gamma$ and $v_{\rm C\,\textsc{iv}}$ are correlated to the growth rate history of the SMBH. Although additional data are needed to confirm it, this result may suggest that, in luminous $z>6$ QSOs, the SMBH predominantly grows via fast accretion rather than via initial high seed BH mass.}
\maketitle
%
\nolinenumbers
\section{Introduction}
\label{sect:intro}
The study of luminous ($L_{\rm bol}\gtrsim10^{47}$ erg\,s$^{-1}$) quasars (QSOs) at $z>6$, hosting supermassive black holes (SMBHs) with black hole masses $M_{\rm BH} > 10^8 M_{\odot}$ up to $10^{10} M_{\odot}$ 
\citep[][and references therein]{2010A&ARv..18..279V,2011ApJ...736...28W,2015Natur.518..512W,2016PASA...33....7J,2017PASA...34...31V,2018ApJ...856L..25B,2020ApJ...897L..14Y,Wang2021a,2023ApJ...950...68E,2023ARA&A..61..373F,Zappacosta2023,2023MNRAS.523.1399D}, offers a unique window to investigate their formation and rapid growth in the short time interval available ($<1$\,Gyr). The formation of SMBH on a relatively short timescale is still an open question and, assuming the formation of high-$z$ SMBH progenitors at $z\approx 20-30$ \citep{2016PASA...33....7J}, different scenarios have been presented to explain the presence of a $10^9M_{\odot}$ SMBH at $z>6$. They either involve the presence of a massive BH seed ($M^{\rm seed}_{\rm BH}>10^{3-4}M_{\odot}$) regardless of the subsequent accretion rate and/or a series of short and intermittent super-Eddington accretion phases allowing the growth from lower BH mass seeds ($M^{\rm seed}_{\rm BH}\sim100M_{\odot}$) \citep{2016MNRAS.456.2993L,2020ARA&A..58...27I}. Currently there are no conclusive indications towards one scenario or the other. If, on the one hand, the existence of a local population of active intermediate-mass BH (10$^{4.5-5.2} M_{\odot}$) in the local Universe \citep{Reines2013,Chilingarian2018,Greene2020} may support stellar mass BH seeds for their origin \citep{Mezcua2017}, on the other hand, the observation at $z\gtrsim10$ of BHs with $M_{\rm BH} \gtrsim 10^6 M_{\odot}$, direct progenitors of the first ($z>6$) luminous QSOs, still cannot provide conclusive evidence for their origin \citep{2024Natur.630E...2M,2024NatAs...8..126B}.

Information about the accretion process of SMBHs can be obtained by exploring the innermost regions of luminous QSOs, for example through X-ray spectroscopy. X-ray emission arising from SMBHs, powering active galactic nuclei (AGN) and luminous QSOs, in particular, is believed to be produced by the interplay between the accretion disc (AD) and the corona, called the 'two-phase model' \citep{Haardt1991}. Thermal UV/optical photons emitted from the AD are inverse-Compton scattered by the coronal hot relativistic electrons into the X-rays, creating a primary X-ray continuum (e.g. \citealt{1980A&A....86..121S}; \citealt{1993ApJ...413..507H}) with the spectral shape of a power law characterized by a photon index, $\Gamma$, and a cut-off at high energy, $E_{\rm cut}$, both related to the physical characteristics of the corona (i.e. coronal temperature, $kT_e$, and optical depth, $\tau$). This high-energy radiation is a direct manifestation of the extreme conditions near the central SMBH, and it carries essential information about the innermost regions of the AGN, the mechanisms governing their accretion process and ultimately the SMBH growth. The photon index of the primary power law is a possible proxy of the AGN accretion rate, parametrized by the Eddington ratio: the ratio of the bolometric luminosity to the Eddington luminosity, $\lambda_{\rm Edd}=\frac{L_{\rm bol}}{L_{\rm Edd}}$. However, the presence of a $\Gamma-\lambda_{\rm Edd}$ relation, according to which very steep $\Gamma$ are commonly detected in highly accreting AGN, has been largely debated \citep{2017MNRAS.470..800T,2021ApJ...910..103L,2022A&A...657A..57L,2022ApJ...927...42K,2023A&A...677A.111T,2023MNRAS.519.6267T}.

There are many observational works dedicated to the X-ray spectroscopy of $z>6$ QSOs so far \citep[e.g.][and references therein]{2017MNRAS.467.3590G,2018ApJ...856L..25B,2019A&A...631A.120S,2019MNRAS.484.5142P,Vito2019,Connor2019,Connor2020,Vito2021,Wang2021b,2022ApJ...924L..25Y,Zappacosta2023}. In particular, \citet[][]{Zappacosta2023} reports the result of the X-ray analysis of the first year of a \xmm Multi--Year Heritage programme dedicated to the HYPerluminous quasars at the Epoch of ReionizatION, (HYPERION) sample, which consists of 18 $z>6$ luminous ($L_{\rm bol}>10^{47}$\,erg\,s$^{-1}$) QSOs, known by 2020, and powered by SMBHs that appear to have undergone the fastest SMBH formation compared to other coeval sources. Assuming continuous exponential growth via accretion at $\lambda_{\rm Edd}=1$, these QSOs require an initial $M^{\rm seed}_{\rm BH}>1000M_{\odot}$, assuming that the formation of their seeds happend at $z=20$, e.g. \citet{2016MNRAS.457.3356V}. \citet{Zappacosta2023} found that the X-ray photon index is, on average, significantly steeper than that of $z<6$ QSOs that are analogues in terms of luminosity and $\lambda_{\rm Edd}$ therefore suggesting a redshift evolution in the nuclear properties of the first QSOs \citep[see also][]{Vito2019,Wang2021a}.

Luminous QSOs are also expected to show the presence of powerful winds at all scales \citep[e.g.][]{2012MNRAS.425..605F,2014Natur.513..210S,2017A&A...601A.143F}. In particular, AD winds traced by broad C\,\textsc{iv} emission lines have been discovered with velocities, corresponding to the relative shift between the peak of the C\,\textsc{iv} and Mg\,\textsc{ii} emission lines, up to $v_{\rm C\,\textsc{iv}}\sim -8000$\,km/s \citep{2016ApJ...831....7S}. The $v_{\rm C\,\textsc{iv}}$ parameter is found to correlate with the QSOs accretion rate \citep{Richards_2011,2016Ap&SS.361...29M,2018A&A...617A..81V,2020MNRAS.492.4553R,2020MNRAS.492..719T,2021MNRAS.501.3061T}. Notably, \citet{Zappacosta2020} reported an anti-correlation between $v_{\rm C\,\textsc{iv}}$ and the intrinsic 2--10\,keV luminosity ($L_{\rm x}$) for a sample of luminous ($L_{\rm bol}>10^{47}$ erg\,s$^{-1}$) $z=2-4$ QSOs with similar UV luminosity. This suggests a connection between the AD winds and the X-ray emission whereby the stronger winds are hosted in weaker X-ray sources.

For this work we explored for the first time the relations between the X-ray nuclear properties (i.e. $\Gamma$ and $L_{\rm x}$), the C\,\textsc{iv} velocity shift, the C\,\textsc{iv} rest-frame equivalent width ($REW_{\rm C\,\textsc{iv}}$) and properties regarding the physics and growth of the AD and the SMBH, in a sample of luminous QSOs at $z\sim6-7.5$. The paper is organized as follows. In Section \ref{sect:sample_and_data_red}, we present the sample and we describe the data reduction processes. In Section \ref{sect:Analysis} we present the spectral and correlation analysis performed in this work. In Section \ref{sect:results} we describe the results of our analysis, which are discussed in Section \ref{sect:discussion} and summarized in Section \ref{sect:conclusion}.

Standard cosmological parameters (H=70\,km\,s$^{-1}$\,Mpc$^{-1}$, $\Omega_{\Lambda}$=0.73 and $\Omega_m$=0.27) are adopted throughout the paper. Errors are reported at the 68\% confidence level with upper and/or lower limits quoted at the 90\% confidence level.

\section{Sample selection and data presentation}
\label{sect:sample_and_data_red}
\subsection{The sample}
\label{subsec:sample}

\begin{table*}
\caption{Sample of $z>6$, luminous QSOs considered in this work, along with their accretion and C\,\textsc{iv} emission properties.}
\label{table:sources}  
\centering 
\begin{tabular}{lcccccccc}   
\hline\hline                
Target &  $z^a$& $\log(L_{\rm bol}^b)$ & log($M_{\rm BH}^c)$& $\lambda_{\rm Edd}$ &$\log(M^c_{\rm s, Edd})$ & $v_{\rm C\,\textsc{iv}}^d$ & $REW^d_{\rm C\,\textsc{iv}}$ & Ref.  \\
       &   & [erg\,s$^{-1}$]       & [$M_{\odot}$]      &                     & [$M_{\odot}$]      &   [km\,s$^{-1}$]               &[$\AA$] &      \\  
\hline
\noalign{\smallskip}
\multicolumn{9} {c}{HYPERION QSOs: \xmm Heritage programme} \\
\noalign{\smallskip}
ULAS\,J1342+0928 & 7.541& $47.19\pm 0.01$& $8.90\pm0.14$& 1.58 & $4.28\pm0.12$& $-5633\pm828$ & $17.04\pm1.04$  & 1 \\
J1007+2115       & 7.494& $47.30\pm0.02$ & $9.18\pm0.05$& 1.06 & $4.51\pm0.05$&$-3201\pm918$  & $10.00\pm1.60$  & 1 \\
ULAS\,J1120+0641 & 7.087& $47.30\pm0.21$ & $9.41\pm0.11$& 0.61 & $4.26\pm0.10$&$-2276\pm183$  & $31.05\pm1.70$  & 1 \\
DES\,J0252-0503  & 6.99 & $47.12\pm0.04$ & $9.15\pm0.05$& 0.74 & $3.88\pm0.05$&$-4354\pm762$  & $17.30\pm1.00$  & 1 \\
VDES\,J0020-3653 & 6.834& $47.16\pm0.01$ & $9.24\pm0.08$& 0.66 & $3.75\pm0.09$&$-1700\pm100$  & $55.00\pm1.00$  &  1 \\
VHS\,J0411-0907  & 6.824& $47.31\pm0.02$ & $8.80\pm0.04$& 2.57 & $3.30\pm0.04$&$-1418\pm298$  & $43.20\pm4.80$  & 1 \\
VDES\,J0244-5008 & 6.724& $47.19\pm0.01$ & $9.08\pm0.15$& 1.02 & $3.45\pm0.18$&$-3200\pm310$  & $24.00\pm2.00$  & 1 \\
PSO\,J231.6-20.8 & 6.587& $47.31\pm0.01$ & $9.50\pm0.09$& 0.51 & $3.67\pm0.08$&$-3829\pm116$  & $6.66\pm2.00$   &  1, 5 \\
PSO J036.5+03.0  & 6.533& $47.33\pm0.05$ & $9.49\pm0.12$& 0.55 & $3.58\pm0.14$&$-4477\pm326$  & $20.78\pm0.90$  &  1, 5 \\
VDES\,J0224-4711 & 6.526& $47.53\pm0.01$ & $9.36\pm0.08$& 1.18 & $3.43\pm0.09$&$-1814\pm258$  & $50.25\pm2.00$  &  1, 5 \\
PSO\,J011+09     & 6.444& $47.12\pm0.01$ & $9.15\pm0.15$& 0.74 & $3.10\pm0.18$&$-3356\pm338$  & $7.38\pm1.75$   &  1, 6 \\
SDSS\,J1148+5251 & 6.422& $47.57\pm0.01$ & $9.74\pm0.03$& 0.54 & $3.66\pm0.001$& $-2803\pm51$   & $44.68\pm2.46$  &  1, 6 \\
SDSS\,J0100+2802 & 6.300& $48.24\pm0.04$ &$10.04\pm0.27$& 1.26 & $3.76\pm0.04$& $-2496\pm316$  & $5.11\pm0.50$   &   1, 5 \\
ATLAS\,J025-33   & 6.294& $47.39^{*}$ & $9.43\pm0.21$&  0.73 & $3.14\pm0.27$&$-3246\pm295$  & $18.82\pm1.50$  &   1, 8 \\
CFHQS\,J0050+3445& 6.246& $47.29\pm0.01$ & $9.68^{*}$&  0.32 & $3.31^{*}$&$864\pm487$    & $63.68\pm2.60$  & 1 \\
ATLAS\,J029-36   & 6.027& $47.39^{*}$ & $9.82^{*}$& 0.30 & $3.08^{*}$& $-2178\pm267$  & $14.59\pm1.90$  & 1, 8 \\
\noalign{\smallskip}
\hline
\noalign{\smallskip}
\multicolumn{9} {c}{Non-HYPERION QSOs} \\
\noalign{\smallskip}
PSO\,J159-02     & 6.38$^{\dagger}$ & $47.29\pm0.01$ & $9.51\pm0.05$ & 0.30 & $3.32\pm0.05$  &$-726\pm120$  & $54.70\pm3.82$  & 4, 6, 7 \\
SDSS\,J1030+0524 & 6.308& $47.14\pm0.14$ & $9.29\pm0.10$ & 0.57 & $3.02\pm0.11$  &$-876\pm235$   & $32.77\pm2.37$ & 2, 5 \\
PSO\,J308-21     & 6.24$^{\dagger}$ & $47.37\pm0.01$ &$9.24\pm0.07$ & 1.40 & $2.83\pm0.08$   &$-2003\pm233$  & $35.15\pm1.61$ &  3, 6 \\
SDSS\,J1602+4228 & 6.09$^{\star}$ & $47.03^{*}$ & $9.37^{*}$ & 0.62 & $2.57^{*}$  &$-311\pm479$   & $59.19\pm3.23$ & 2, 5 \\
SDSSJ1306+0356   & 6.034$^{\dagger}$& $47.12\pm0.01$ & $9.31\pm0.07$ & 0.48 &$2.59\pm0.12$ &$-786\pm111$    & $47.89\pm1.76$ &  2, 5 \\
\hline\hline
\end{tabular}
\tablefoot{$^a$:$z$ estimated from the Mg\,\textsc{ii} emission line if not stated otherwise;  $^b$: estimated from luminosity 3000$\AA$ \citep{2006ApJS..166..470R}; $^c$: for the HYPERION sources we report the values of the $M_{\rm BH}$ and the estimated $M_{\rm s, Edd}$ from \citet{Zappacosta2023}, for the Non-HYPERION from \citet{Mazzucchelli2023,Farina2022} re-scaled using the same cosmological estimators as in \citet{Zappacosta2023}, if present; $^d$: When more than one $v_{\rm C\,\textsc{iv}}$ and $REW_{\rm C\,\textsc{iv}}$ values is present in the literature, for simplicity, we report the mean values derived using all the available values reported in Table \ref{table:vshift}, the errors are the mean of the errors. We note that the the correlation analysis takes into account all the existing values of $v_{C\,\textsc{iv}}$ (see Section \ref{subsect:fitting}).\\
$^{\dagger}$: $z$ estimated from the [C\,\textsc{ii}] emission line.\\
$^{\star}$: $z$ estimated from the Ly\,$\alpha$ emission line.\\
$^{*}$: Error not reported in the literature.}\\
\tablebib{(1)~\citet{Zappacosta2023}; (2)~\citet{Vito2019}; (3)~\citet{Connor2019}; (4)~\citet{Pons2020}; (5)~\citet{Mazzucchelli2023}; (6)~\citet{Farina2022} (7)~\citet{Schindler2020}; (8)~\citet{Chehade2018}.}
\end{table*}

The sample considered in this work includes all known $z>6$, $L_{\rm bol}\gtrsim10^{47}$\,erg\,s$^{-1}$ QSOs detected with $>30$ total X-ray net counts (0.3--7\,keV) and with available measurements of $v_{\rm C\,\textsc{iv}}$ from the literature. It consists of 16 HYPERION QSOs with X-ray data from the \xmm\ Multi--Year Heritage programme obtained by August 2023 and five other QSOs with available archival good--quality X-ray data from \citet{Connor2019},\citet{Vito2019} and \citet{Pons2020}. 
The list of all 21 QSOs and their general properties is reported in Table \ref{table:sources}. We included in the analysis the most recent \xmm\ data of the HYPERION source ATLAS\,J029-36 (OBSID 0930591101 and 0930591201, P.I. Norbert Schartel), which helped us to improve considerably the data quality of the HYPERION dataset.

For our QSO sample we introduced the seed mass parameter assuming accretion at the Eddington limit ($\lambda_{\rm Edd}=1$), \mseed, defined using the exponential relation
\begin{equation}
M_{\rm s, Edd}=M_{\rm BH} \times e^{-t/t_s}
\label{eq:mseed}
\end{equation}
where $t$ is the elapsed time between the formation redshift of the seed BH and the redshift at which the QSO is observed and $t_s$ is the e-folding time: $t_s=0.45\eta(1-\eta)^{-1}\lambda_{\rm Edd}^{-1}f_{\rm duty}^{-1}$\,[Gyr].
In this calculation we assumed that seed BHs form at $z=20$ \citep{2016MNRAS.457.3356V} and continuously accrete at the Eddington limit, with radiative efficiency, representing the fraction of accreted mass which is radiated, $\eta=0.1$, and with duty cycle (i.e. the fraction of the time during which the AGN is active, $f_{\rm duty}=1$) corresponding to a continuous active phase. This quantity can be considered as a proxy for the SMBH growth rate: the larger \mseed, the higher the expected growth rate.

We note that the HYPERION QSOs have BH masses computed from single-epoch Mg\textsc{ii}-based estimation \citep{2009ApJ...699..800V}. For some of the HYPERION QSOs (i.e. VHS\,J0411-0907 and VDES\,J0020-3653 \citep{2023A&A...678A.191M}; VDES\,J0244-5008 and VDES\,J0224-4711 \citep{2023ApJ...951L...5Y}; SDSS\,J0100+2802 \citep{2023ApJ...950...68E}) H\,$\beta$-based BH masses estimates from the \textit{James Webb} Space Telescope (JWST) are now available. However, the H\,$\beta$-based BH masses are consistent with Mg\textsc{ii}-based masses, ensuring statistically the reliability of \mseed\ estimates.

Within the \xmm Multi-Year Heritage programme, most of the sources have observations during multiple epochs. We consistently reduced and analysed all the \xmm data of the 16 HYPERION QSOs, as well as the available \xmm data of PSO\,J159-02 and SDSS\,J1030+0524, two of the five additional non-HYPERION sources in our sample. The X-ray data reduction and their spectral analysis are described in Section \ref{subsect:data_red} and Section \ref{subsect:Spectral_analysis}, respectively. For the remaining three QSOs, which only have {\it Chandra} observations available, we used data from the literature. 

Regarding $v_{\rm C\,\textsc{iv}}$ and $REW_{\rm C\,\textsc{iv}}$, there are multiple values in the literature of both parameters, measured with different fitting approaches (see Section \ref{subsect:vshift} and Table \ref{table:vshift}). In this work, we took into account all the available measures in the literature of $v_{C\,\textsc{iv}}$ and/or $REW_{\rm C\,\textsc{iv}}$ (see Section \ref{subsect:fitting} for more details).

\subsection{X-ray data reduction}
\label{subsect:data_red}
\begin{table*}
\caption{\xmm\ observations analysed in this work.}
\label{table:obsid}  
\centering 
\begin{tabular}{llcccrrrr}   
\hline\hline            
OBSID      & Target          & RA          &DEC          &Start date                &Exp. & \multicolumn{3} {c}{Net Exp.}\\
           &                 &(J2000)      & (J2000)     & (UTC)                    & [ks]& PN [ks] & MOS1 [ks] & MOS2 [ks]  \\
\hline
\noalign{\smallskip}
\multicolumn{9} {c}{HYPERION QSOs: archival observations} \\
\noalign{\smallskip}
0790180701 & SDSS\,J0100+2802 & 01:00:13.02  & +28:02:25.8 & 2016-06-29 & 65.00 & 44.85 & 59.92 & 56.00\\
0824400301 & VDES\,J0224-4711 & 02:24:26.54  & -47:11:29.4 & 2018-05-25 & 35.00 & 13.87 & 27.81 & 24.73\\
0930591101 & ATLAS\,J029-36   & 01:59:57.97  &-36:33:56.6 & 2024-06-10  & 135.00 & 61.79 & 75.03 & 78.02 \\
0930591201 & ATLAS\,J029-36   & 01:59:57.97  &-36:33:56.6 & 2024-06-12  & 135.00 & 67.98 & 86.07 & 91.11\\
\noalign{\smallskip}
\hline
\noalign{\smallskip}
\multicolumn{9} {c}{HYPERION QSOs: \xmm Heritage programme QSOs} \\
\noalign{\smallskip}
0884992601 & CFHQS\,J0050+3445& 00:50:06.67  & +34:45:22.6 & 2021-06-26    & 45.00 & 26.38 & 37.39 & 29.66\\
0884990401 & ULAS\,J1120+0641 & 11:20:01.48  & +06:41:24.3 & 2021-06-27    & 73.00 & 36.08 & 56.27 & 55.38\\
0884990101 & ULAS\,J1342+0928 & 13:42:08.10  & +09:28:38.6 & 2021-07-05    & 106.50 & 60.51 & 82.48 & 68.38\\
0884992101 & PSO\,J011+09     & 00:45:33.57  &+09:01:56.9  & 2021-07-15    & 81.00 & 45.36 & 58.97 & 48.78\\
0884992001 & PSO\,J036.5+03.0 & 02:26:01.88  &+03:02:59.4  & 2021-07-19    & 85.00 & 47.56 & 68.69 & 57.24\\
0884991701 & PSO\,J231.6-20.8 & 15:26:37.83  & -20:50:00.7 & 2021-07-29    & 109.00 & 67.07 & 90.45 & 79.05\\
0884991501 & VDES\,J0244-5008 & 02:44:01.02  & -50:08:53.7 & 2021-08-04    & 89.00 & 67.23 & 82.87 & 73.48\\
0884993801 & ULAS\,J1342+0928 & 13:42:08.10  & +09:28:38.6 & 2021-12-24    & 102.00 & 46.53 & 55.36 & 54.09\\
0884991101 & VDES\,J0020-3653 & 00:20:31.47  & -36:53:41.8 & 2022-01-01    & 87.00 & 36.54 & 64.67 & 61.26\\
0884992901 & ATLAS\,J029-36   & 01:59:57.97  &-36:33:56.6  & 2022-01-03    & 85.00 & 55.65 & 71.49 & 65.45\\
0886201201 & J1007+2115        & 10:07:58.26 & +21:15:29.2 & 2022-05-28    & 83.40 & 51.37 & 67.37 & 66.06\\
0886210301 & VHS\,J0411-0907   & 04:11:28.62 & -09:07:49.7 & 2022-07-31    & 95.05& 53.12 & 81.62 & 78.07\\
0886210801 & PSO\,J231.6-20.8  & 15:26:37.83 & -20:50:00.7 & 2022-08-14    &103.00 & 52.19 & 78.10 & 36.97\\
0886220301 & SDSS\,J1148+5251  & 11:48:16.64 & +52:51:50.2 & 2022-11-08   & 86.70 & 55.77 & 68.70 & 67.78\\
0886210201 & VDES\,J0020-3653  & 00:20:31.46 & -36:53:41.8 & 2022-11-16    & 87.20 &36.90 & 49.05 & 48.24\\
0886220201 & PSO\,J011+09      & 00:45:33.56 & +09:01:56.9 & 2023-01-13    & 88.00 & 50.97 & 69.09 & 68.32\\
0884990901 & DES\,J0252-0503   & 02:52:16.64 & -05:03:31.8 & 2023-01-15    & 87.50 & 27.68 & 35.85 & 36.72\\
0886220701 & CFHQS\,J0050+3445 & 00:50:06.67 & +34:45:22.6 & 2023-02-01    & 34.90 & 4.78 & 12.21 & 12.04\\
0886200201 & J1007+2115        & 10:07:58.26 & +21:15:29.2 & 2023-05-14    & 83.80 & 56.89 & 68.05 & 66.32\\
0886221501 & PSO\,J011+09      &00:45:33.56  & +09:01:56.9 & 2023-07-01    & 85.10 & 44.98 & 62.28 & 47.18\\
0886221401 & ATLAS\,J025-33    & 01:42:43.69 & -33:27:45.6 & 2023-07-19    & 94.85 & 43.63 & 81.28 & 60.77\\
0886201001 & DES\,J0252-0503   & 02:52:16.64 & -05:03:31.8 & 2023-07-25    & 89.50 & 37.08 & 50.80 & 49.31\\
0886200901 & DES\,J0252-0503   & 02:52:16.64 & -05:03:31.8 & 2023-07-29    & 83.00 & 68.97 & 80.17 & 78.16\\
0886210901 & PSO\,J231.6-20.8  & 15:26:37.83 & -20:50:00.7 & 2023-07-31    & 86.30 & 36.97 & 59.94 & 58.47\\
\noalign{\smallskip}
\hline
\noalign{\smallskip}
 \multicolumn{9} {c}{Non-HYPERION QSOs}  \\
\noalign{\smallskip}
0803161101 & PSO\,J159-02     & 10:36:54.19 & -02:32:37.94 & 2017-12-22 & 23.00 & 16.97 & 20.87 & 19.81\\
0148560501 & SDSS\,J1030+0524 & 10:30:27.11 & +05:24:55.06 & 2003-05-22  & 103.88 & 51.99 & 64.90 & 66.54\\
\hline\hline
\end{tabular}
\tablefoot{We report the observations of the HYPERION QSOs from the archive and from the \xmm Multi-Year Heritage X-ray programme, up to August 2023, we included also the most recent \xmm\ data of the HYPERION source ATLAS\,J029-36 (OBSID 0930591101 and 0930591201, P.I. Norbert Schartel) and archival \xmm\ observations of the non-HYPERION QSOs.}
\end{table*}
We reported the details of the HYPERION observations, both archival and from the \xmm Multi-Year Heritage X-ray programme (up to August 2023) in the upper part of Table \ref{table:obsid}. The available \xmm observations of the non-HYPERION QSOs belonging to our sample are reported in the lower part of Table \ref{table:obsid}.

The data reduction was performed following the same approach as \citet{Zappacosta2023}. The event lists of the EPIC-pn \citep{Struder2001} and EPIC-MOS \citep{Turner2001} detectors are extracted with the \texttt{epproc} and \texttt{emproc} tools of the standard System Analysis Software (\texttt{SAS} v.18.0.0; \citealt{Gabriel2004}). The latest calibration files available by August 2023 were used. The choice of the optimal time cuts for flaring particle background was performed visually inspecting light curves created in the 10--12\,keV (EPIC-pn) and >10\,keV (EPIC-MOS) energy ranges, with PATTERN=0 (single events). For the choice of the source and background extraction radii, we identified the point-like sources in each target field of view running the meta-task \texttt{edetect-chain} on the $0.5-2$\,keV energy band EPIC images by setting a detection maximum likelihood threshold DETML=6 (DETML=$-\ln P_{\rm rnd}$ where $P_{\rm rnd}$ is the probability of detection by chance). We selected as source region a circular region with a radius of 20 arcsec (corresponding to $\sim 80\%$ of the on-axis PSF encircled energy fraction at 1.5 keV), centred on each published optical target position. Since VDES\,J0244-5008 and VDES\,J0020-3653 had a nearby source distant 28 arcsec and 17 arcsec, respectively, we adopted a smaller source region of 15 arcsec and 12 arcsec radius ($\sim 65\% - 70\%$ of the encircled energy fraction), respectively. As background regions, we adopted for the EPIC-pn camera rectangular regions with long and short sides in the range 3.6--3.9 arcmin and 1.9--2.7 arcmin, respectively, located around the source region and rotated with the same position angle of the detector. For the two EPIC-MOS detectors, we adopted as background regions circular regions of radius in the range 2.5--3.4 arcmin centred on the target position. From all the background regions we excluded circular regions with a radius of 40 arcsec centred on the target position and on all the contaminant point sources previously identified. Response matrices and auxiliary response files were generated using the \texttt{SAS} tasks \texttt{rmfgen} and \texttt{arfgen}, respectively. Spectral data were binned using the optimal \citealt{2016A&A...587A.151K} (hereafter KB) grouping, which provides the optimal binning for data and model accounting for the source spectral shape, the variable spectral resolution and the average photon energy in each bin. The KB grouping is the best scheme to recover unbiased energy independent spectral parameter estimates for low count regime spectra of $z>6$ sources (see Appendix B in \citealt{Zappacosta2023} for further details).

\subsection{C\,\textsc{iv} velocity shift}
\label{subsect:vshift}
\begin{table*}
\caption{Values of $v_{\rm C\,\textsc{iv}}$ and $REW_{\rm C\,\textsc{iv}}$ reported in literature for the sources in our sample.}
\label{table:vshift}  
\centering 
\begin{tabular}{lccc||lccc}   
\hline\hline                
Target &  $v_{\rm C\,\textsc{iv}}$ & $REW_{\rm C\,\textsc{iv}}$  & Ref.&Target &  $v_{\rm C\,\textsc{iv}}$ & $REW_{\rm C\,\textsc{iv}}$  & Ref.\\
& [km\,s$^{-1}$] &[$\AA$]& & & [km\,s$^{-1}$] &[$\AA$]  \\  
\hline  
\noalign{\smallskip}
ULAS\,J1342+0928 &$-5986$& - & 1& PSO\,J036.5+03.0  &$-5386\pm689$ &  $41.5\pm1.1$ &6\\
                 &$-4935\pm758$ &1$2.90\pm1.40$&2& &$-4382$       & - &1\\
                 &$-5978\pm889$ & - & 3& &$-4382$       & - &1\\
                 &  - &$21.18\pm0.69$ & 4& &$-3727\pm135$  & $21.83^{+0.80}_{-1.10}$ & 9, 12\\
\hline 
\noalign{\smallskip}
J1007+2115       &$-3183\pm1475$&$10.00\pm1.60$ & 2 & ATLAS\,J025-33   &$-2461\pm251$&$18.82^{+1.50}_{-1.20}$&9, 12  \\
                 &$-3220\pm362$ &  - & 5 &                  & $-4032\pm340$&  - & 11\\
\hline
\noalign{\smallskip}
DES\,J0252-0503  &$-4618\pm762$ & $17.30\pm1.00$ & 2 & ATLAS\,J029-36   &$-1924\pm377$&$14.59^{+1.90}_{-0.50}$&9, 12 \\
                 &$-4090$ & -& 7 &                  &$-2433\pm158$ & - & 11\\
\hline
\noalign{\smallskip}
VDES\,J0020-3653 &$-1700\pm100$ & $55.00\pm1.00$ & 8 & VDES\,J0244-5008 &$-3200\pm310$&$24.00\pm2.00$ & 8 \\
\hline
\noalign{\smallskip}
PSO\,J231.6-20.8 &$-5131$ & - & 1 & VDES\,J0224-4711 &$-2000\pm160$ & $44.00\pm2.00$ & 8\\
                 &$-2528\pm116$ & $6.66^{+2.00}_{-1.30}$ & 9, 12 &                  &$-1634\pm56$ & - & 3\\
                 &$-5861\pm318$ &$23\pm1.2$&6 & &$-1808\pm42$ &$56.50^{+1.70}_{-1.80}$& 9, 12\\
\hline
\noalign{\smallskip}
PSO\,J011+09     &$-3356\pm338$ & $7.38\pm1.75$ & 4 & SDSS\,J1148+5251 &$-2803\pm51$&$ 44.68\pm2.46$ & 10 \\
\hline
\noalign{\smallskip}
SDSS\,J0100+2802 &$-2496\pm316$&$5.11\pm0.50$&9, 12 & CFHQS\,J0050+3445 &$864\pm487$ &$63.68\pm2.60$&10 \\
\hline
\noalign{\smallskip}
PSO\,J159-02     & $-726\pm120$&$54.70\pm3.82$&4 & PSO\,J308-21     &$-2003\pm233$&$33.15\pm1.61$&4 \\
\hline
\noalign{\smallskip}
SDSS\,J1030+0524 &$-822$&-&1 & SDSSJ\,1306+0356  &$-855$&-&1 \\
                 &$-768\pm379$ &$32.77\pm2.37$&4 & &$-735\pm33$&$47.89\pm1.76$&4\\
                 &$-1092\pm92$ & -&9 & &$-769\pm189$&-&9\\
\hline
\noalign{\smallskip}
ULAS\,J1120+0641 &$-2602\pm285$ & - & 6 & SDSS\,J1602+4228  &$-311\pm479$&$56.19\pm3.23$&10\\
                 &$-2966$ &  - & 1 \\
                 &$-2007\pm133$ & $25.90\pm2.40$  & 2 \\
                 &$-2520\pm199$ & - & 3 \\
                 &$-1583\pm115$ & $33.10\pm1.00$ & 4 \\
\hline\hline
\end{tabular}
\tablebib{(1)~\citet{Meyer2019}; (2)~\citet{Yang2021}; (3)~\citet{Wang2021b}; (4)~\citet{Farina2022};(5)~\citet{2020ApJ...897L..14Y}; (6)~\citet{2017ApJ...849...91M}; (7)~\citet{2020ApJ...896...23W}; (8)~\citet{Reed2019}; (9)~\citet{Mazzucchelli2023}; (10)~\citet{Shen2019}; (11)~\citet{2018MNRAS.478.1649C}; (12)~Bischetti et al., in prep.}
\end{table*}
The shift in the peaks of high-ionization broad emission lines, such as C\,\textsc{iv}, relative to the source systemic redshift, is usually interpreted as a signature of AD-driven broad-line winds \citep{2000ApJ...545...63E,2004ApJ...611..125L} at both low redshifts \citep[e.g.][]{2002AJ....124....1R} and high redshifts \citep[e.g.][]{2014ApJ...790..145D}.

The presence of C\,\textsc{iv} shifts in luminous QSOs with respect to the systemic redshift determined by lower ionization broad lines (e.g. Mg\,\textsc{ii}) or host narrow lines (e.g. [O\,\textsc{iii}], CO, [C\,\textsc{ii}]) has been widely investigated, and the position of the C\,\textsc{iv} line has been determined using the line velocity centroid \citep[e.g.][]{2011Natur.474..616M,10.1093/mnras/stw1360,Reed2019,2020A&A...635A.157T,Yang2021} and/or the peak of a Gaussian or multi-Gaussian profile \citep[e.g.][]{2014ApJ...790..145D,2017ApJ...849...91M,2018A&A...617A..81V,Shen2019,Meyer2019, Schindler2020,Wang2021b,Mazzucchelli2023}. These different $v_{C\,\textsc{iv}}$ definitions, coupled with different fitting approaches for the line modelling involving one or more Gaussians (depending on the signal-to-noise ratio and the resolving power of the analysed spectra) and the pseudo-continuum models of Fe\,\textsc{ii} templates, result in a range of different C\,\textsc{iv} velocity shift values, even for the same source, that differ by up to a factor of 2 (see e.g. ULAS\,J1120+0641 and PSO\,J231.6-20.8). We show in Table \ref{table:vshift} all the values of the C\,\textsc{iv} velocity shift and its rest-frame equivalent width ($REW_{\rm C\,\textsc{iv}}$) reported in the literature to date for each QSO considered in this work. The $REW_{\rm C\,\textsc{iv}}$ of the QSOs belonging to the ESO Large Programme XQR-30 were calculated using the same approach of \citet{Mazzucchelli2023}; details will be reported in Bischetti et al. (in prep). For the analysis performed in this work we took into account multiple values of $v_{C\,\textsc{iv}}$ and/or $REW_{\rm C\,\textsc{iv}}$ (as explained in Section \ref{subsect:fitting}).

\section{Analysis}
\label{sect:Analysis}
\subsection{Spectral analysis}
\label{subsect:Spectral_analysis}
\begin{table*}
\caption{Best-fit parameters from the X-ray spectral analysis.}
\label{table:X-rayparameters}  
\centering 
\begin{tabular}{l c c c c c c c}   
\hline\hline                
Target &  $\Gamma$ & W-stat/dof & $E_{\rm cut}^{a}$ & W-stat/dof &log($L_{\rm x}^{b}$)    & Net counts &Ref. \\
      &            &            & [keV]             &            &[erg\,s$^{-1}$]     &            &\\  
\hline 
\noalign{\smallskip}
\multicolumn{8} {c}{HYPERION QSOs: \xmm Heritage programme}\\
\noalign{\smallskip}
ULAS\,J1342+0928 & $2.93^{+0.43}_{-0.37}$ & 393/319& $8.5_{-5}^{+40}$ & 393/319& $45.22\pm0.08$ & 123$^{\dagger}$ &1 \\
J1007+2115       & $2.66^{+0.78}_{-0.78}$ & 325/309&$14.8_{-10}^{+15}$& 324/309 &$44.50\pm0.09$ & 42$^{\dagger}$ &1 \\
ULAS\,J1120+0641 & $2.65^{+0.35}_{-0.32}$ & 148/155 &$7.5_{-4}^{+20}$ & 148/155 &$45.39\pm0.06$ & 75$^{\dagger}$  &1 \\
DES\,J0252-0503  & $2.63^{+0.92}_{-0.92}$ &368/347 & $3.9_{-2}^{+30}$& 369/347& $45.08\pm0.05$ & 55$^{\dagger}$ & 1 \\
VDES\,J0020-3653 & $2.73^{+0.45}_{-0.32}$ & 273/305& $9.2_{-5}^{+25}$& 273/305&$45.12\pm0.08$ & 80$^{\dagger}$ & 1 \\
VHS\,J0411-0907  & $1.84^{+0.48}_{-0.48}$ & 192/159& $>50$           & 193/159&$45.13\pm0.05$ & 134$^{\dagger}$ & 1 \\
VDES\,J0244-5008 & $2.35^{+0.30}_{-0.25}$ & 298/309 &$>25$           & 298/309&$45.28\pm0.04$ &  124$^{\dagger}$ & 1 \\
PSO\,J231.6-20.8 & $2.30^{+0.62}_{-0.67}$ & 918/841& $>13$           &918/841& $45.09\pm0.09$ & 170$^{\dagger}$  & 1 \\
PSO\,J036.5+03.0 & $3.05^{+1.10}_{-0.85}$ & 64/60  &$10.5_{-5}^{+50}$&65/60& $44.95\pm0.11$ & 52$^{\dagger}$   & 1 \\
VDES\,J0224-4711 & $2.08^{+0.20}_{-0.20}$ & 181/153 &$>28$           &182/153& $45.58\pm0.05$ & 97$^{\dagger}$   & 1 \\
PSO\,J011+09     & $2.58^{+0.82}_{-0.82}$ & 340/310 &$>13$           &339/310 &$45.04\pm0.11$ & 91$^{\dagger}$   & 1 \\
SDSS\,J1148+5251 & $2.38^{+0.39}_{-0.39}$ & 168/157 &$16.4_{-8}^{+60}$&168/157&$45.18\pm0.08$ & 153$^{\dagger}$  & 1 \\
SDSS\,J0100+2802 & $2.41^{+0.15}_{-0.09}$ & 160/165 &$31.0_{-13}^{+53}$&160/165& $45.76\pm0.03$ & 290$^{\dagger}$  & 1 \\
ATLAS\,J025-33   & $2.03^{+0.42}_{-0.37}$ & 166/156 &$>20$            &165/156& $45.24\pm0.03$ & 182$^{\dagger}$  & 1 \\
CFHQS\,J0050+3445& $2.02^{+0.43}_{-0.41}$ & 198/191 &$>50$            &198/191& $45.07\pm0.08$ & 80$^{\dagger}$   & 1 \\
ATLAS\,J029-36   & $2.54^{+0.24}_{-0.21}$ & 492/475& $34_{-10}^{+13}$ &492/475& $44.91\pm0.08$ & 259$^{\dagger}$   & 1 \\
\noalign{\smallskip}
\hline
\noalign{\smallskip}
\multicolumn{8} {c}{Non-HYPERION QSOs}\\
\noalign{\smallskip}
PSO\,J159-02     & $1.86^{+0.45}_{-0.40}$&139/155& $>23$ & 139/155& $45.41\pm0.11$ &  90$^{\dagger}$ & 1 \\
SDSS\,J1030+0524 & $2.19^{+0.20}_{-0.19}$&183/165& $29^{+58}_{-13}$ &182/165& $45.59\pm0.09$ & 207$^{\dagger}$ & 1 \\
PSO\,J308-21     & $2.39^{+0.36}_{-0.37}$& - & - & -&$45.36\pm0.15$ & 72$^{\star}$  & 3 \\
SDSS\,J1602+4228 & $2.19^{+0.74}_{-0.71}$& - & - & -&$45.56\pm0.08$ & 30$^{\star}$  & 2 \\
SDSS\,J1306+0356 & $1.83^{+0.26}_{-0.25}$& - & - &-&$45.23\pm0.05$ & 133$^{\star}$ & 2 \\
\hline\hline
\end{tabular}
\tablefoot{$a$: This cut-off value is obtained when applying the \texttt{zcutoffpl} model in \texttt{xspec} with $\Gamma$ fixed to a value of $1.9$; $b$: $L_{\rm x}$ is measured in the 2--10\,keV rest-frame energy band.\\
$^{\dagger}$: Sum of the EPIC-pn, EPIC-MOS1 and EPIC-MOS2 net counts in the 0.3--7\,keV energy range.\\
$^{\star}$: {\it Chandra} total net counts in the 0.5--7\,keV energy range.}
\tablebib{(1) This work; (2)~\citet{Vito2019}; (3)~\citet{Connor2019}.}
\end{table*}
The spectral analysis was performed using the \texttt{xspec} v.12.12.1 software package \citep{Arnaud1996}. Following \citet{Zappacosta2023}, we performed the modelling by using the Cash statistics with direct background subtraction \citep[W-stat in \texttt{xspec},][]{1979ApJ...228..939C,1979ApJ...230..274W}. We adopted a simple power law model, modified by Galactic absorption only, and included a constant in the model to take into account the possible flux variations of the sources within observations taken at different epochs (parametrized by \texttt{const}*\texttt{tbabs}*\texttt{zpowerlaw} in \texttt{xspec})\footnote{We choose to use a simple power--law model absorbed just by the Galactic column density since all the targets are type 1 sources. Moreover, \citet{Zappacosta2023} verified that the absorption in these QSOs is $N_{\rm H}<10^{21-22}$\,cm$^{-2}$ and, given their high-$z$, the resulting rest--frame region of the spectrum is not sensitive to such column densities.}. For the Galactic column density at the position of the sources, we adopted \citealp{HI4PI2016} maps. The analysis was performed over the optimized 0.3--7\,keV energy range (see Appendix C in \citealt{Zappacosta2023}) which corresponds to a rest-frame energy band spanning from $\sim 2$\,keV to $\sim50$\,keV, modelling the EPIC-pn together with the two EPIC-MOS camera and leaving as free parameters $\Gamma$ and the normalization.

The analysis of the observations performed during the first year of the \xmm Multi-Year Heritage programme are presented in \citet{Zappacosta2023}. We have included these data in the current spectral analysis. No flux and/or spectral variations have been detected in the sources that have observations during multiple epochs. The values obtained from the spectral analysis are reported in Table \ref{table:X-rayparameters}, together with the same parameters taken from the literature for all the sources that were not analysed for this work (i.e. SDSS\,J1602+4228, SDSSJ1306+0356, PSO\,J308-21; see Ref. column in Table \ref{table:X-rayparameters}).

The most remarkable result arising from the spectral analysis of our sample of $z>6$ QSOs is the steep shape of their X-ray spectra \citep[as previously reported by][]{Zappacosta2023}. The derived steep X-ray continuum slopes might be, partly or entirely, associated with a particularly low $E_{\rm cut}$, well below 50 keV. In the presence of limited quality spectra such as those we are reporting for $z>6$ QSOs, steep X-ray continuum slopes can be obtained by either steep $\Gamma$ power laws or by canonical $\Gamma$ power laws (e.g. $\Gamma=1.9$) with a high-energy cut--off $E_{\rm cut}<20$\,keV. Hence, we tested the latter hypothesis on the sources in our sample for which we have \xmm data, by modelling their spectra with a cut--off power law model (\texttt{const}*\texttt{tbabs}*\texttt{zcutoffpl} in \texttt{xspec}), fixing the photon index to $\Gamma=1.9$. We were able to measure the cut-off value for only the 50\% of the spectra; we found lower limits for the others (see Table \ref{table:X-rayparameters}). 

Steep $\Gamma$ and $E_{\rm cut}<20$\,keV are both tracers of relatively cold coronae. The fact that we found generally steep continuum slopes could indicate low coronal temperatures. Our data does not allow us to discriminate between a simple power--law model or a cut--off power--law model, and thus for the following correlation analysis we use the better constrained $\Gamma$ to characterize the X-ray continuum.

\subsection{Correlation analysis}
\label{subsect:fitting}
We checked the relations between the coronal X-ray properties, the velocities of the AD winds and other parameters regarding the physics of the AD and the growth of the SMBH of the QSOs in our sample. The parameters related to the coronal properties used in this analysis are $\Gamma$ and $L_{\rm x}$. As tracers of the AD winds, we used $v_{\rm C\,\textsc{iv}}$ and $REW_{\rm C\,\textsc{iv}}$. We also tested the relation of the coronal X-ray properties with some physical properties of the QSOs in the sample, such as $L_{\rm bol}$, $\lambda_{\rm Edd}$ and $M_{\rm BH}$. For the SMBH growth properties the redshift, $z$, and \mseed\ are used.

The presence of a possible relation between two quantities was investigated by using the \texttt{linmix} code, a hierarchical Bayesian model for fitting a straight line to data with errors in the x and the y directions \citep{Kelly_2007}. To perform the fitting we used a linear model to the data using the relation
\begin{equation}
    y=\rm{A}x+\rm{B}
    \label{eq:fitting_rel}
\end{equation}
where x is the independent variable.

For more than half of the sample sources, multiple values of $v_{\rm C\,\textsc{iv}}$ and/or $REW_{\rm C\,\textsc{iv}}$ are reported in the literature, often measured with different methods (see Section \ref{subsect:vshift}). To take this into account, we iterated the fitting process 10,000 times, each time randomly choosing one value to consider among those available for the QSOs showing more than one $v_{\rm C\,\textsc{iv}}$/$REW_{\rm C\,\textsc{iv}}$ value. The final values of slope, intercept (with relative uncertainties), correlation coefficient and null hypothesis probabilities are the mean among all the values of the fits within $\pm1\sigma$ from the peak of the $\chi^2$ distribution.

Table\,\ref{table:all_fitting_param} reports the best-fit slopes and intercepts along with the significance of the correlation analysis and correlation coefficients for the relations investigated in this work.
\section{Results}
\label{sect:results}
\begin{figure}
\centering
\includegraphics[width=1.\columnwidth]{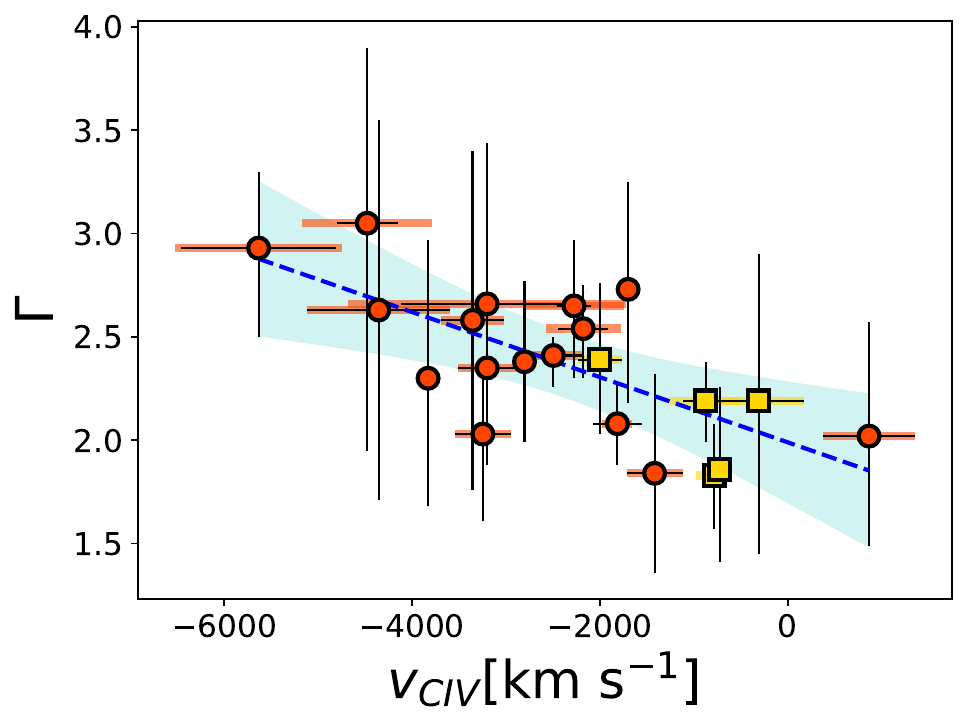}
\caption{Plot of $\Gamma$ vs. $v_{\rm C\,\textsc{iv}}$. The orange circles and the yellow squares represent the HYPERION and non-HYPERION QSOs, respectively. For simplicity we report in the plot the mean values of $v_{C\,\textsc{iv}}$ (see Table \ref{table:sources}), but the fit takes into account all the existing values of $v_{C\,\textsc{iv}}$ (see Section \ref{subsect:fitting}). The thick coloured errorbars represent the maximum deviation from the mean, while the thin black errorbars represent the mean error. The dashed blue line is the linear regression while the shaded region represents the combined 1$\sigma$ error on the slope and normalization (see Table\,\ref{table:all_fitting_param} ). This is valid for all the following plots.}
\label{fig:main_results}
\end{figure}
\begin{figure*}
 \centering
 \includegraphics[width=0.8\columnwidth]{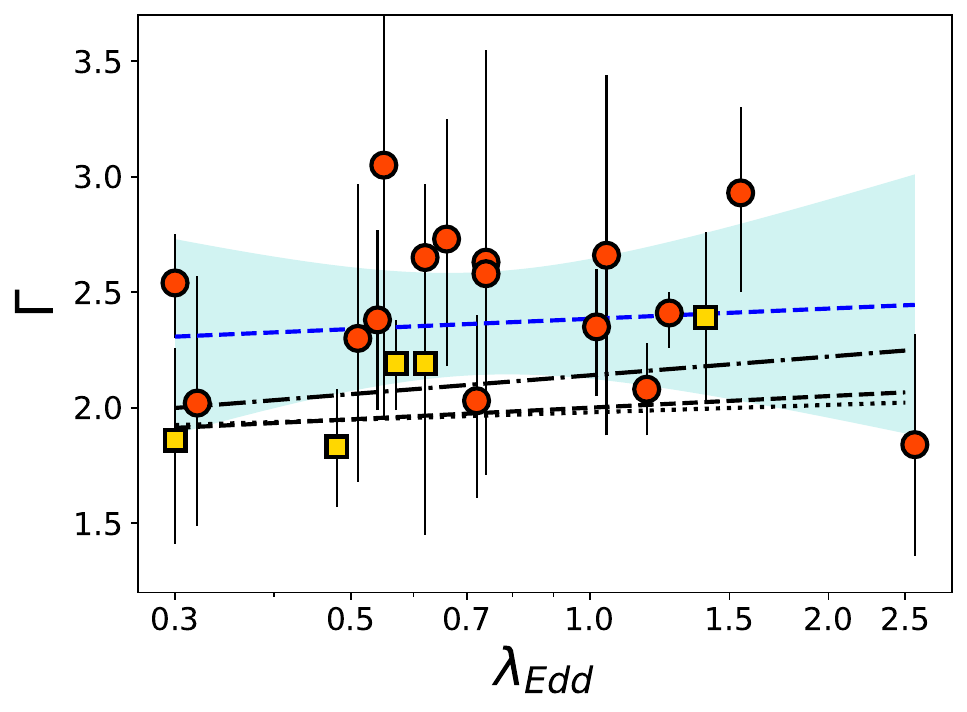}
 \includegraphics[width=0.8\columnwidth]{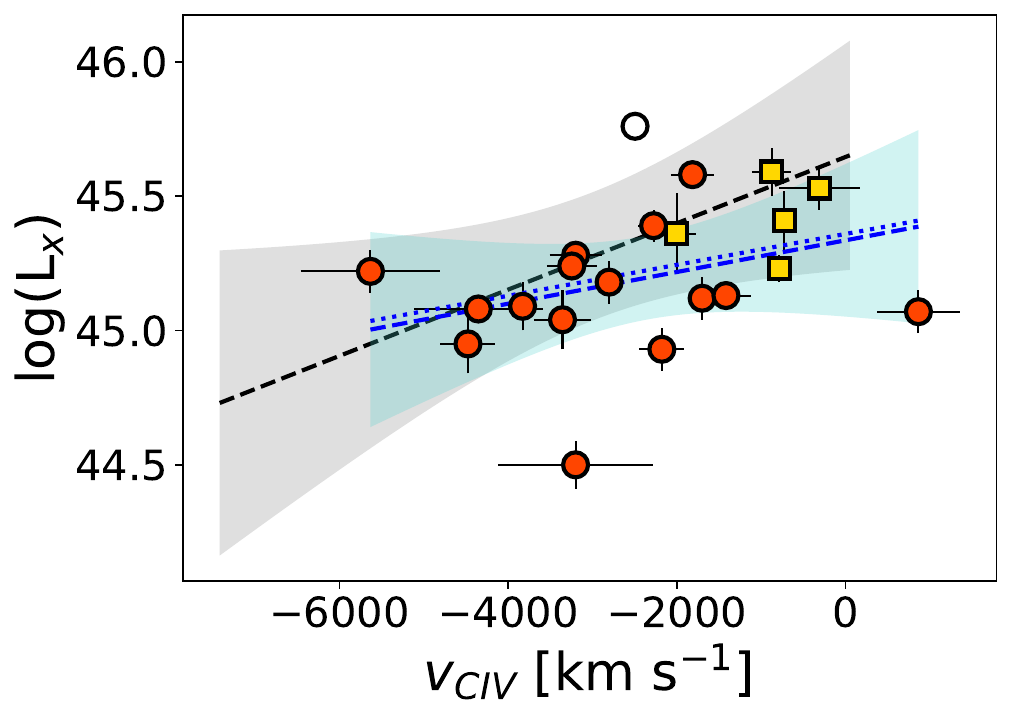}
  \includegraphics[width=0.8\columnwidth]{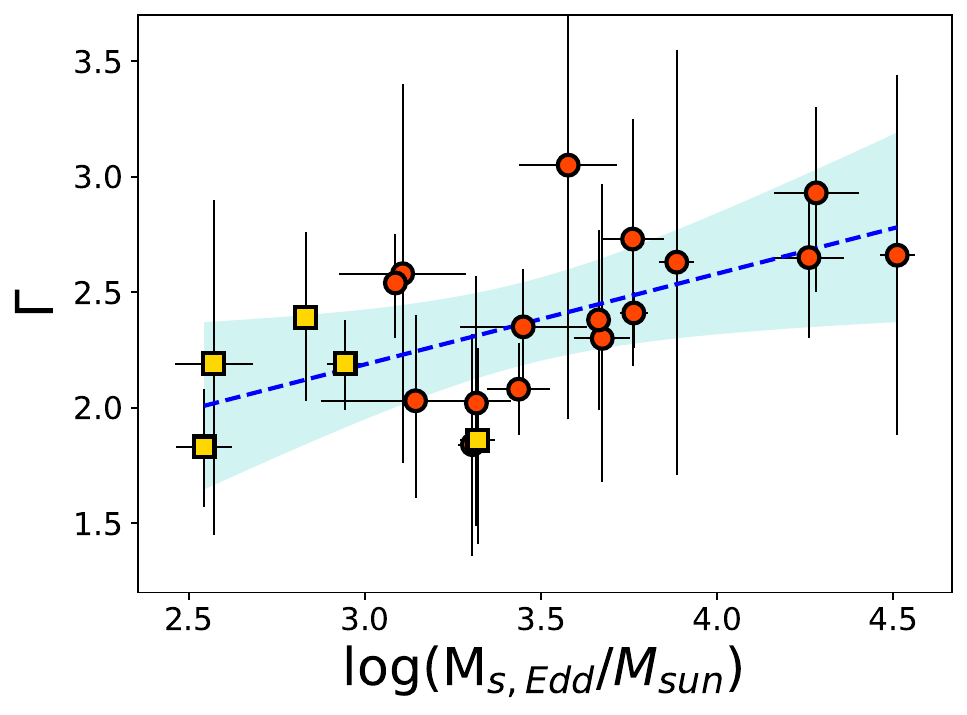}
 \includegraphics[width=0.8\columnwidth]{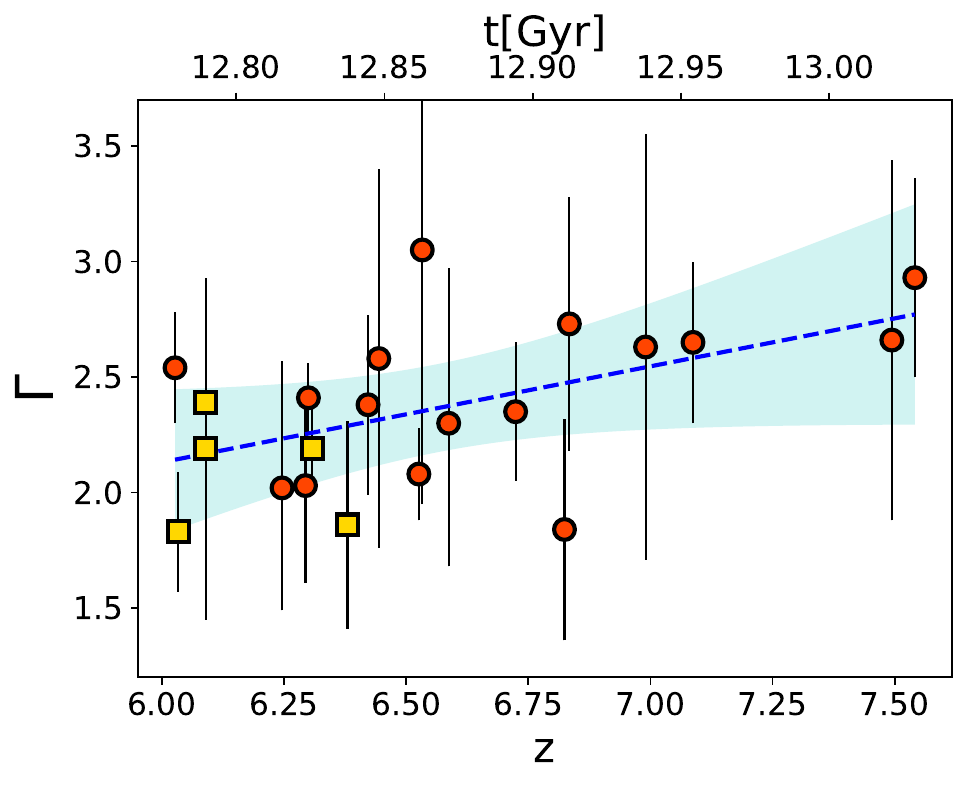}
 \caption{Plots of some of the correlations tested for this work. Upper left panel: $\Gamma$ vs. $\log(\lambda_{\rm Edd})$. We report for comparison the linear relations from \citealt{2017MNRAS.470..800T} (black dashed line), \citealt{2021ApJ...910..103L} (black dot-dashed line) and \citealt{2023MNRAS.519.6267T} (black dotted line); Upper right panel: $v_{C\,\textsc{iv}}$ vs. $L_{\rm x}$, we also report the linear relation (black dashed line) and the combined 1$\sigma$ error on the slope and normalization (grey shaded region) from \citet{Zappacosta2020}. The empty dot is SDSS\,J0100+2802, which is not included in the fitting process given its different value of $L_{\rm bol}$ (see Section \ref{subsect:testing_accr_properties}). The blue dotted line is the linear relation when including SDSS\,J0100+2802. We also report the linear relation from \citealt{Zappacosta2020} (black dotted line). The $v_{\rm C\,\textsc{iv}}$ values plotted here are, for simplicity, the mean values derived using all the available values reported in Table \ref{table:vshift}. We note that the fit takes into account all the existing values of $v_{C\,\textsc{iv}}$ (see Section \ref{subsect:fitting}); Lower left panel: $\Gamma$ vs. \logmseed; Lower right panel: $\Gamma$ vs. $z$.}
   \label{fig:other_relation}
\end{figure*}
\begin{figure}
\centering
\includegraphics[width=0.9\columnwidth]{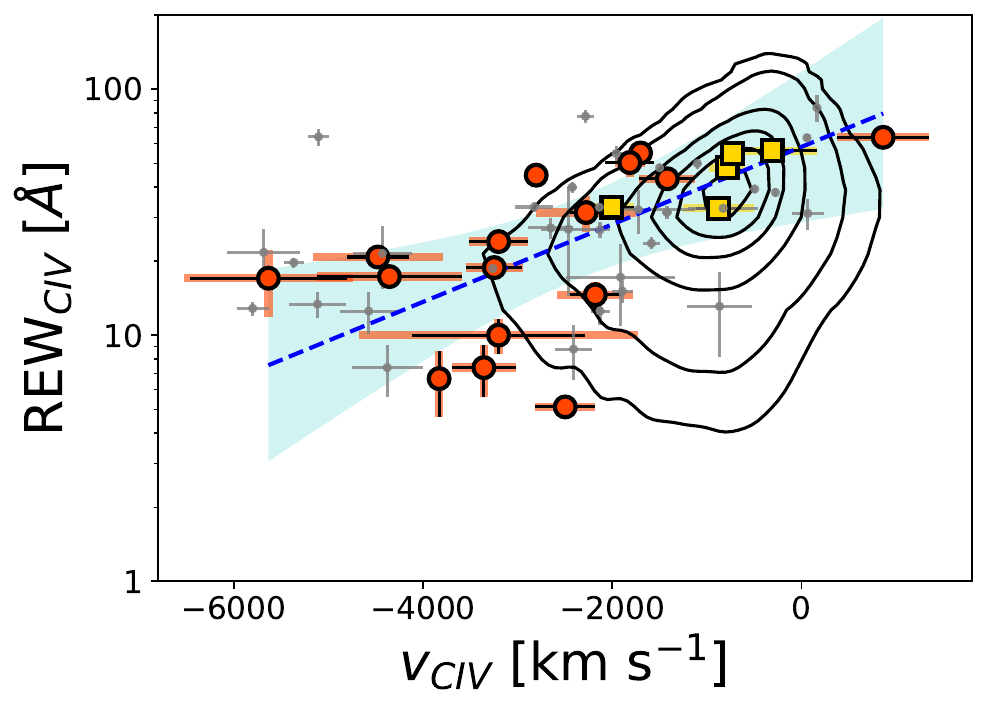}
\caption{The $REW_{\rm C\,\textsc{iv}}$ vs. $v_{\rm C\,\textsc{iv}}$ relation compared with the 0.05, 0.1, 0.3, 0.5, 0.68, 0.9 and 0.99 contour levels (relative to the peak) of the SDSS DR7 sample \citep{2011ApJS..194...45S,2018A&A...617A..81V} and with the values from \citealt{Schindler2020} (grey points). For simplicity, we plot here the mean values derived using all the available values reported in Table \ref{table:vshift}. We note that the fit takes into account all the existing values of $v_{C\,\textsc{iv}}$ and $REW_{\rm C\,\textsc{iv}}$(see Section \ref{subsect:fitting}).}
\label{fig:main_results2}
\end{figure}
Among the relations involving X-ray parameters, we found a significant ($>3\sigma$) relation between $\Gamma$ and $v_{\rm C\,\textsc{iv}}$ (see Figure \ref{fig:main_results}). To compute this relation we used the method explained in Section \ref{subsect:fitting}, which allowed us to take into account all the different values of $v_{\rm C\,\textsc{iv}}$ reported in the literature (see also Section \ref{subsect:vshift}). We checked the presence of the $\Gamma - \lambda_{\rm Edd}$ relation in the QSOs in our sample but we did not find any significant correlation (see Table\,\ref{table:all_fitting_param}  and upper left panel of Figure \ref{fig:other_relation}).

Following the results presented in \citet{Zappacosta2020}, we verified the presence of a relation between $v_{C\,\textsc{iv}}$ and $L_{\rm x}$. We performed this analysis on the sources with similar $L_{\rm bol}$ (within 0.5 dex), removing the obvious outlier of SDSS\,J0100+2802 that exhibits a $L_{\rm bol}$ much larger than other QSOs in the sample. We found a mild relation ($\sim 1.5 \sigma$) between $v_{C\,\textsc{iv}}$ and $L_{\rm x}$ for the presented sample of $z>6$ QSOs. However, the trend is consistent, within the errors, with what was found by \citet{Zappacosta2020} (see upper right panel of Figure \ref{fig:other_relation} and Table\,\ref{table:all_fitting_param} ).

In light of the marginal evidence ($\sim 2.2 \sigma$) for a dependence between \mseed\ and $\Gamma$ reported in \citet{Zappacosta2023}, we tested whether it could be potentially associated with an evolutionary phenomenon or with the particularly fast SMBH mass growth of these sources. Thus, we investigated for a possible relation between $\Gamma$, \mseed\ and $z$. We identified moderate log(\mseed)$-\Gamma$ ($2.5\sigma$) and $z-\Gamma$ ($2.3\sigma$) relations (see lower left and lower right panels of Figure \ref{fig:other_relation} respectively, and Table\,\ref{table:all_fitting_param} ).

A significant relation ($>3\sigma$) is also found between $REW_{\rm C\,\textsc{iv}}$ and $v_{\rm C\,\textsc{iv}}$ (see Figure \ref{fig:main_results2}). This trend is already well established in QSOs \citep{2011AJ....141..167R,2017MNRAS.464.3431H,2018A&A...617A..81V,2020A&A...644A.175V,2021MNRAS.501.3061T,Schindler2020,matthews2023disc}. Assuming the C\,\textsc{iv} line profile is the result of the combination of a standard peaked virialized component and an outflowing shallower component, it is observed that, as the wind velocity increases, the standard virialized component of the line profile decreases. 

\begin{figure*}
   \centering
   \includegraphics[width=0.2\textwidth]{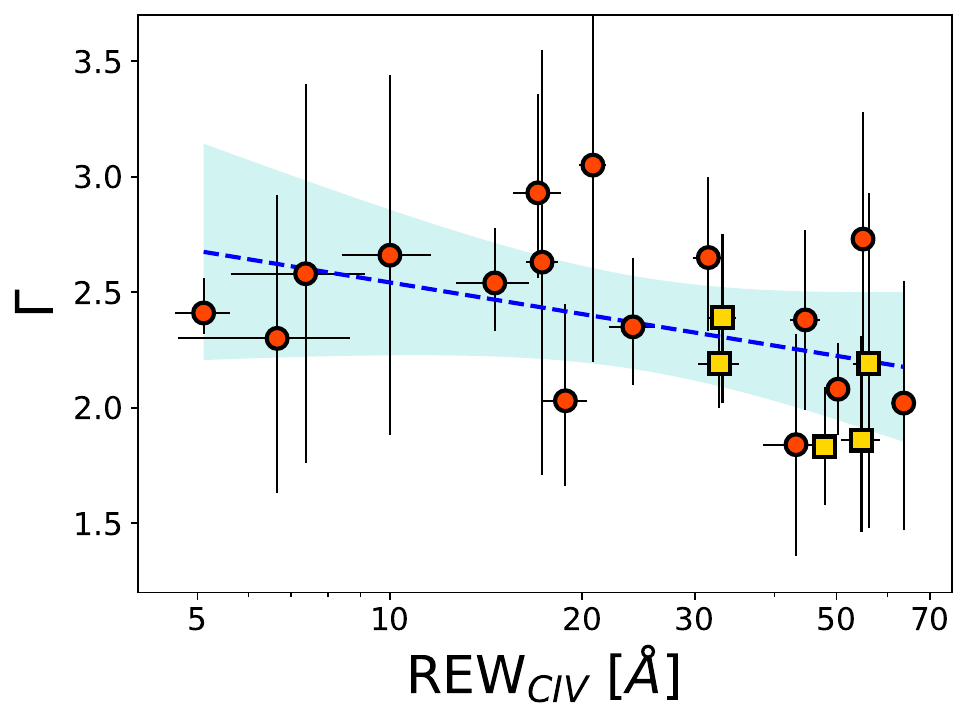}
   \includegraphics[width=0.2\textwidth]{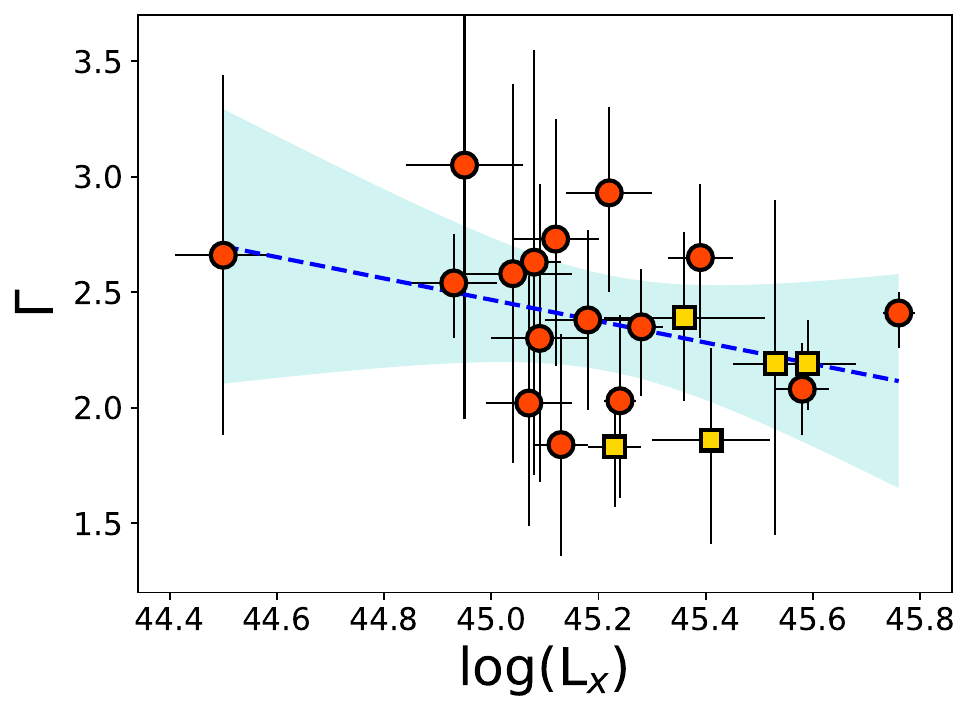}
   \includegraphics[width=0.2\textwidth]{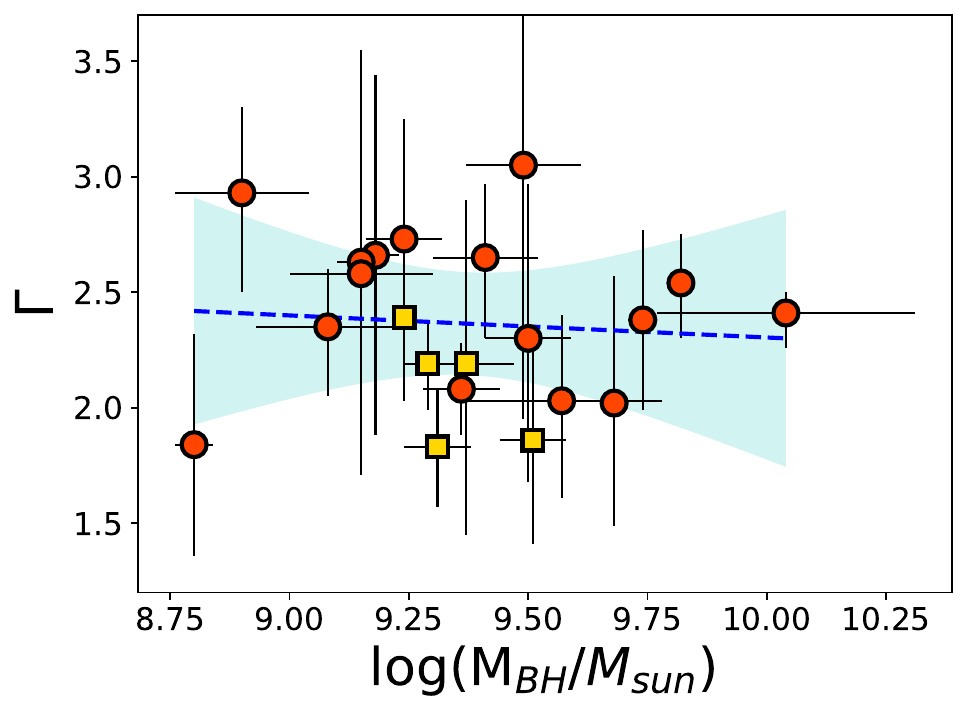}
   \includegraphics[width=0.2\textwidth]{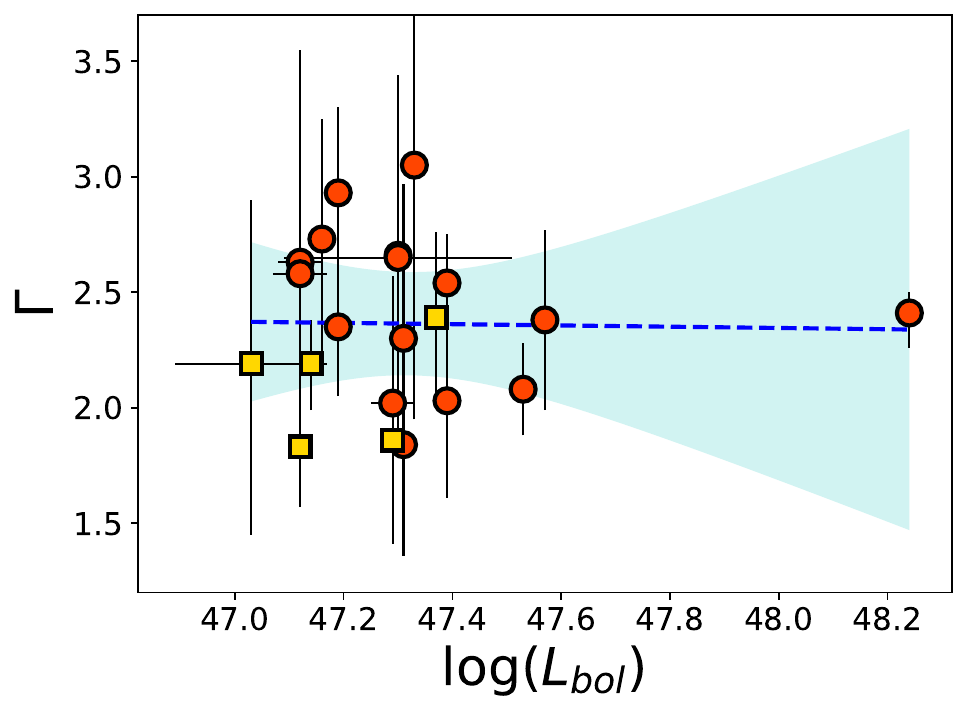}
   \includegraphics[width=0.2\textwidth]{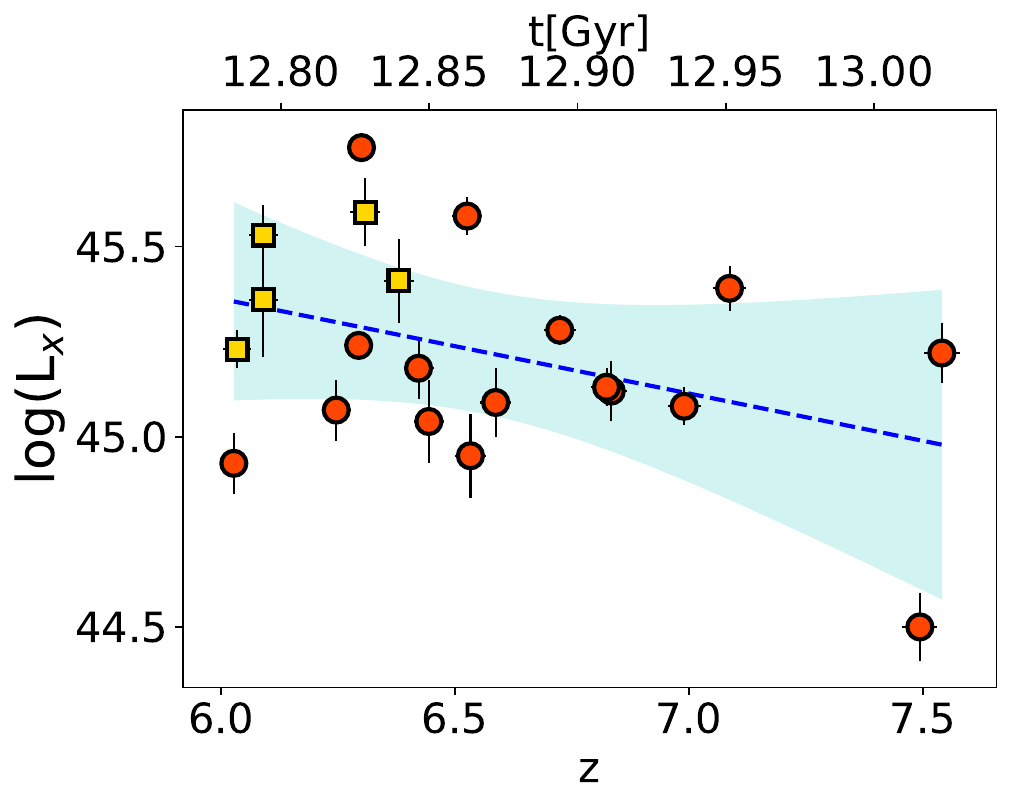}
    \includegraphics[width=0.2\textwidth]{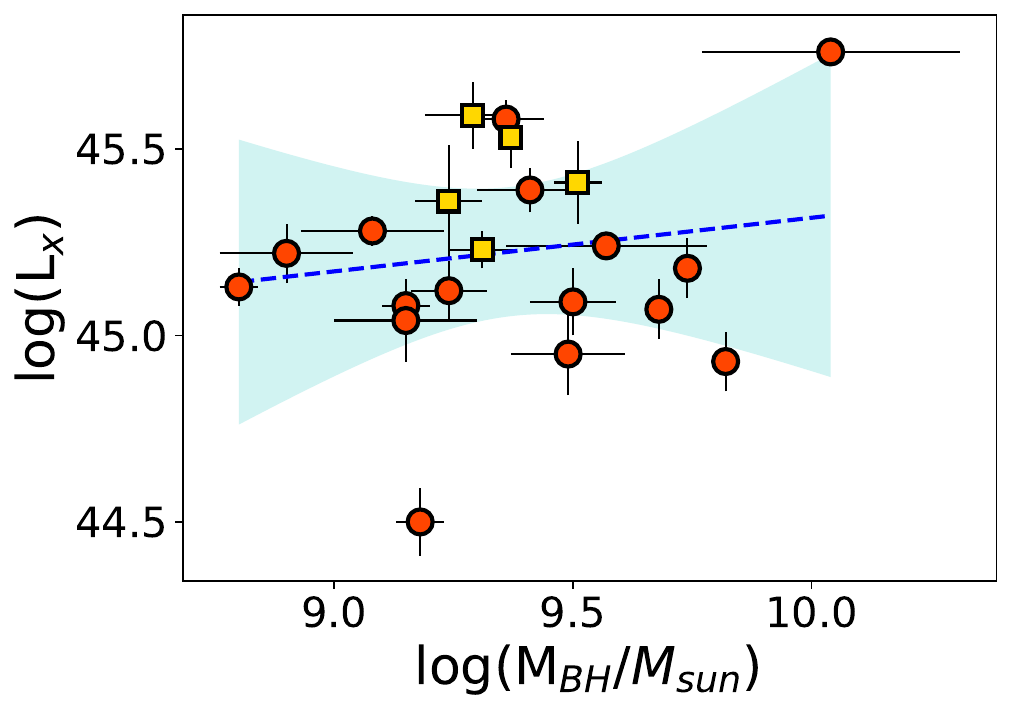}
    \includegraphics[width=0.2\textwidth]{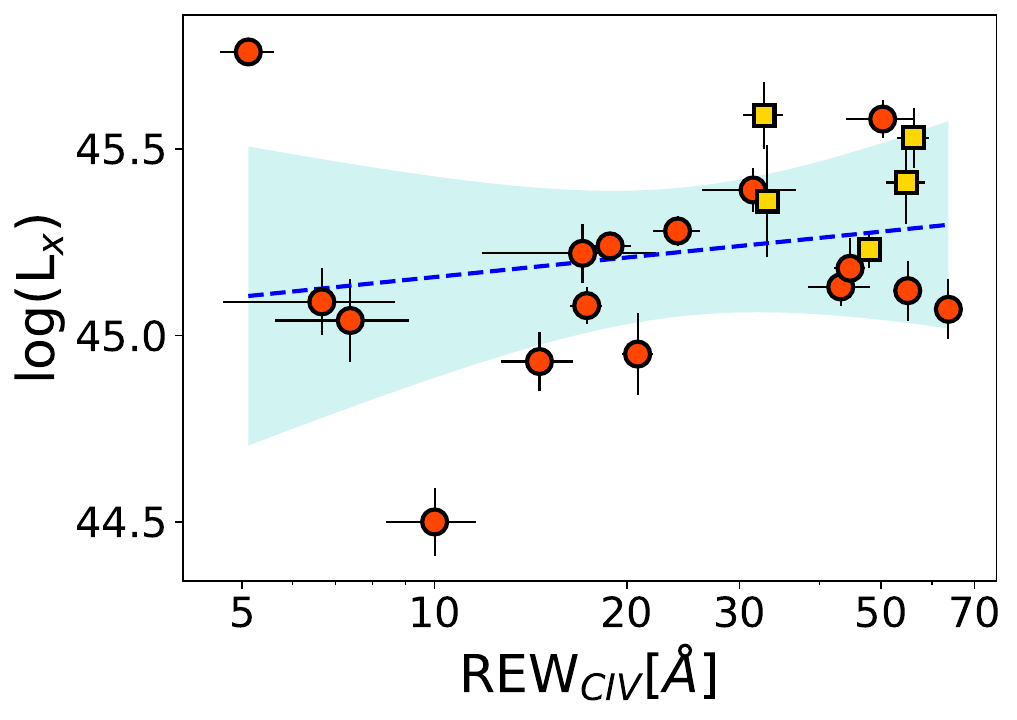}
    \includegraphics[width=0.2\textwidth]{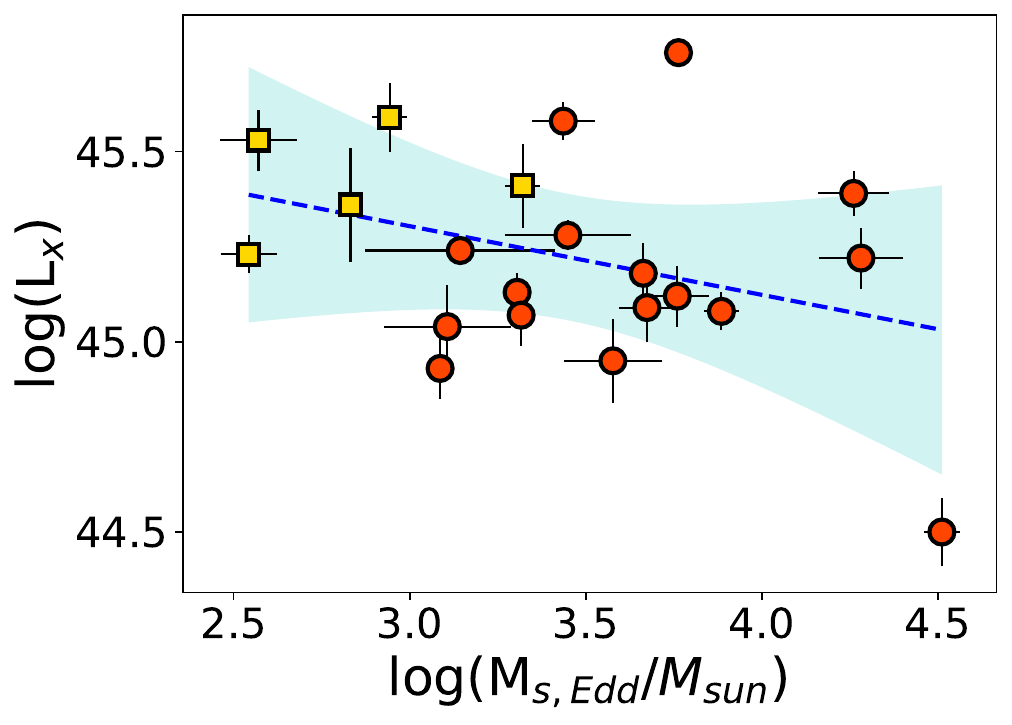}
    \includegraphics[width=0.2\textwidth]{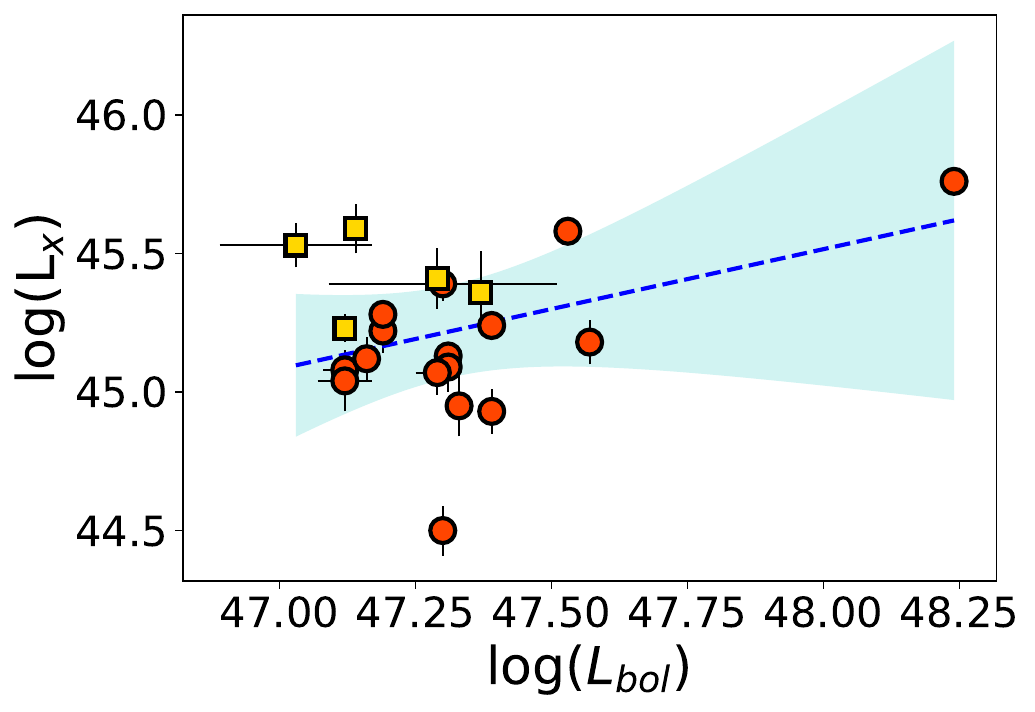}
    \includegraphics[width=0.2\textwidth]{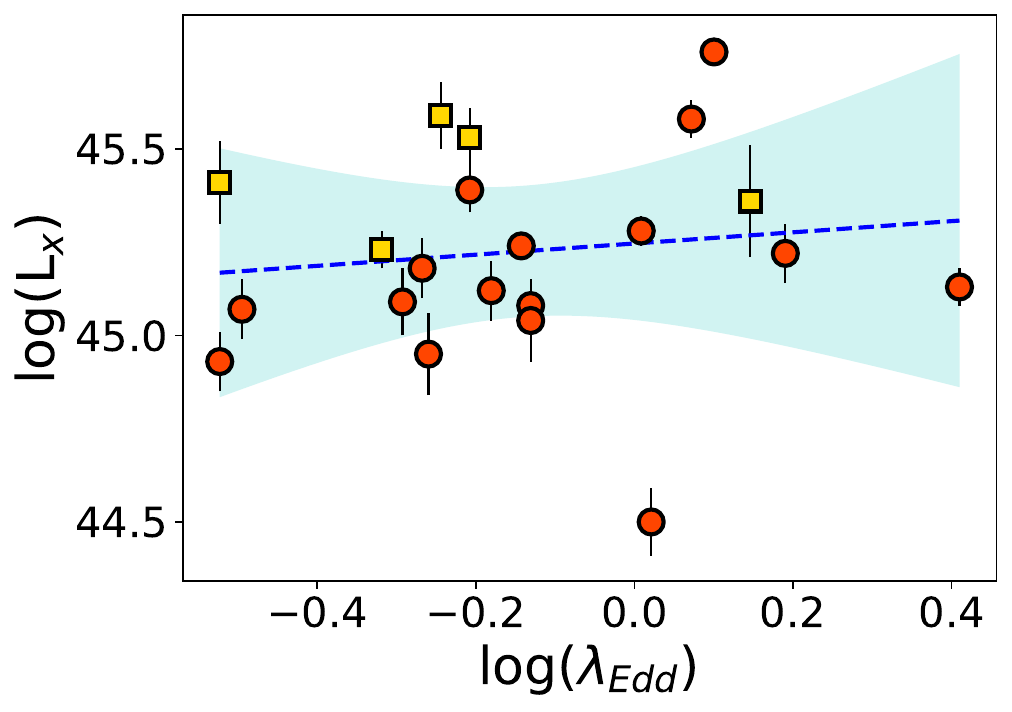}
    \includegraphics[width=0.2\textwidth]{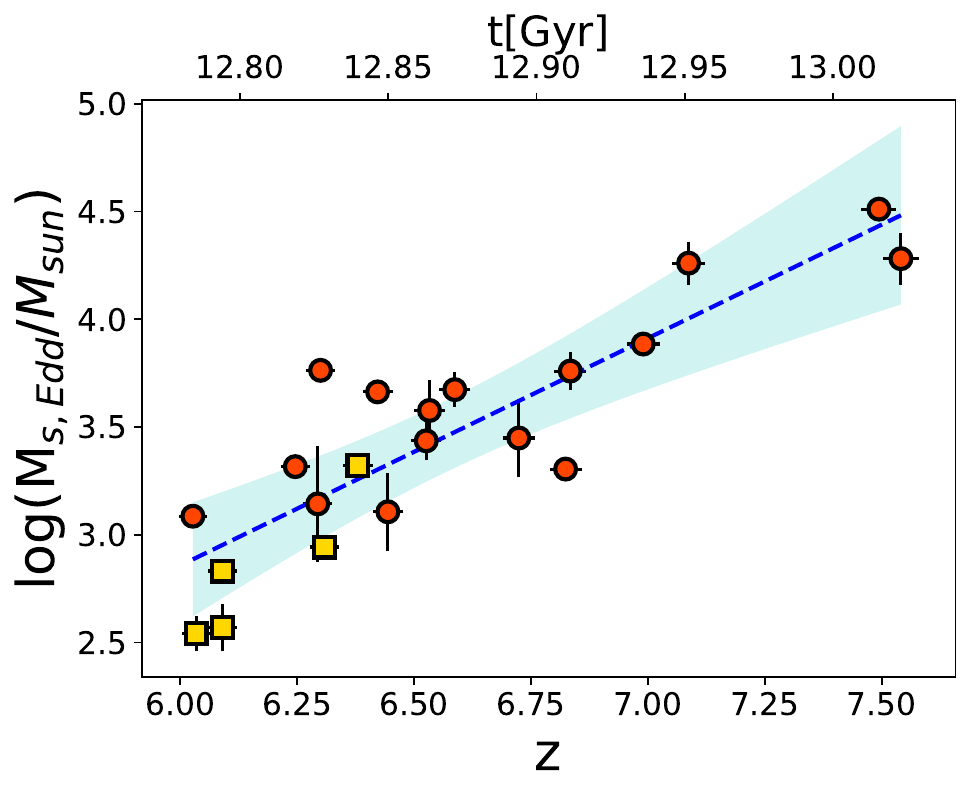}
    \includegraphics[width=0.2\textwidth]{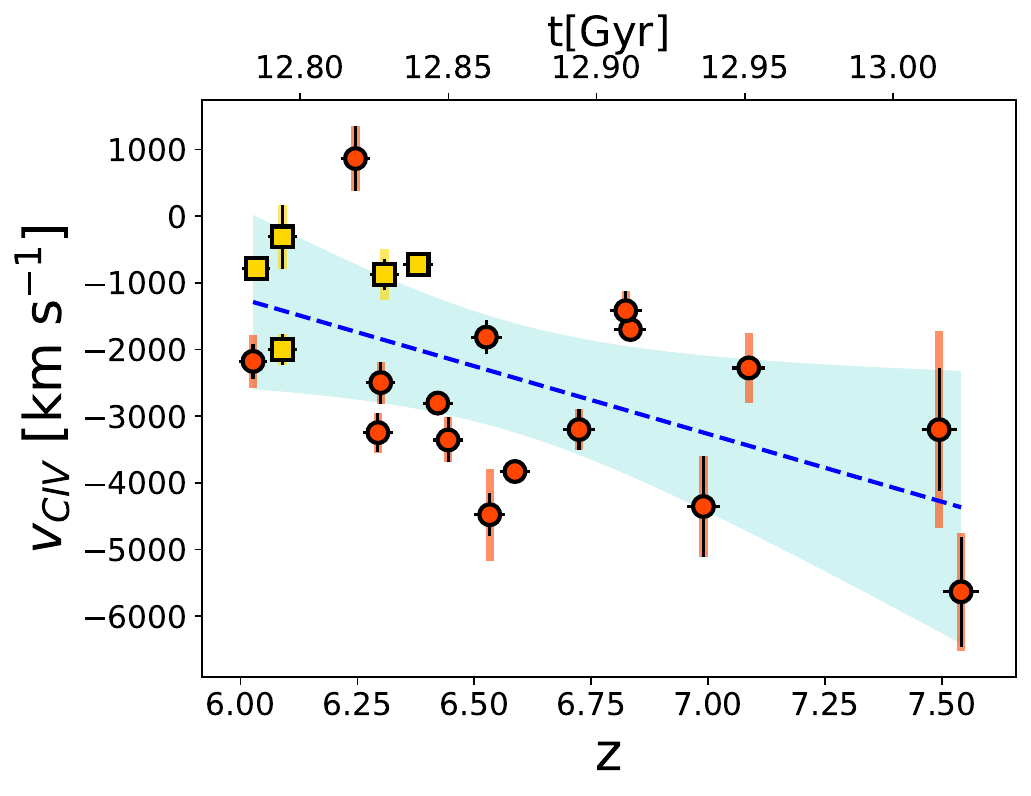}
    \includegraphics[width=0.2\textwidth]{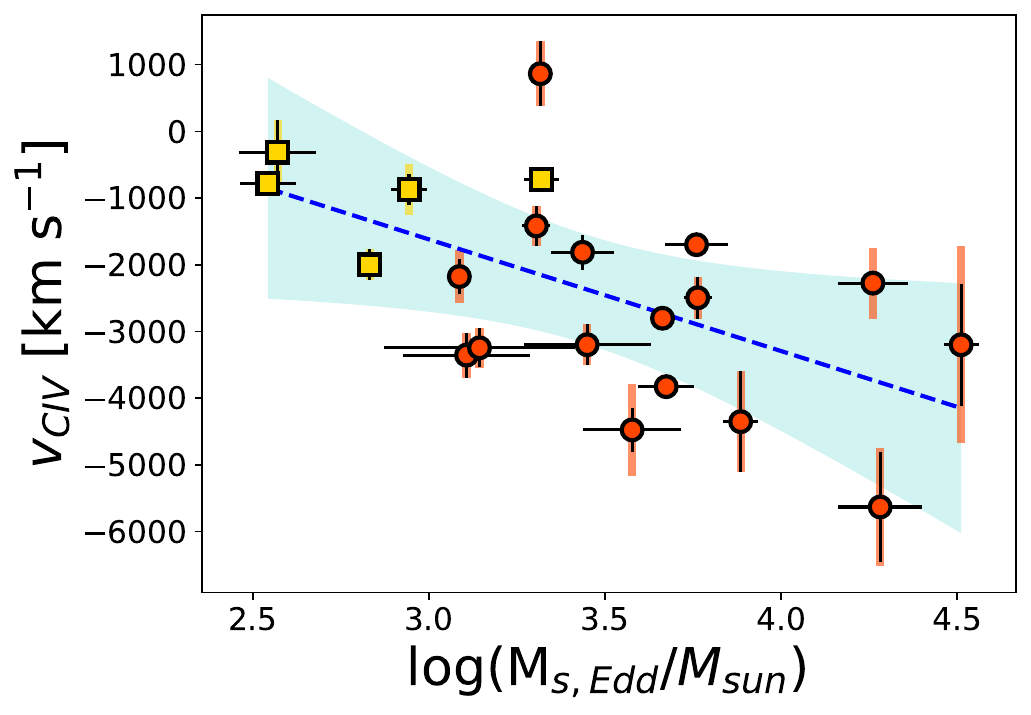}
   \caption{From top to bottom and from left to right, we report: $\Gamma$ vs. $REW_{\rm C\,\textsc{iv}}$; $\Gamma$ vs. $\log(L_{\rm x})$; $\Gamma$ vs. $\log(M_{\rm BH})$; $\Gamma$ vs. $\log(L_{\rm bol})$; $\log(L_{\rm x})$ vs. $z$; $\log(L_{\rm x})$ vs. $\log(M_{\rm BH})$; $\log(L_{\rm x})$ vs. $REW_{\rm C\,\textsc{iv}}$; $\log(L_{\rm x})$ vs. log(\mseed); $\log(L_{\rm x})$ vs. $\log(L_{bol})$; $\log(L_{\rm x})$ vs. $\Gamma$; $\log(L_{\rm x})$ vs. $\log(\lambda_{\rm Edd})$; log(\mseed) vs. $z$; $v_{C\,\textsc{iv}}$ vs. $z$;  $v_{C\,\textsc{iv}}$ vs. log(\mseed).}
   \label{fig:other_relation_bis}
\end{figure*}

Other tested relations are also reported in Table\,\ref{table:all_fitting_param}, where we list the best-fitting parameters. The plots are reported in Figure \ref{fig:other_relation_bis}. Among these other tested relations, we note that a relation is found between $REW_{\rm C\,\textsc{iv}}$ and $\Gamma$ with a significance of $\sim 2\sigma$ but this is a by-product of the $v_{\rm C\,\textsc{iv}} - \Gamma$ and $v_{\rm C\,\textsc{iv}} - REW_{\rm C\,\textsc{iv}}$ relations. Moreover,  we report a significant ($>3\sigma$) \mseed$- z$ relation (see Table\,\ref{table:all_fitting_param}); however, it is due to the definition of \mseed\ (see Eq.\,\ref{eq:mseed}) where there is a time (i.e. redshift) dependence.
\begin{table*}
\caption{Best-fit linear relations (y=Ax+B) tested in this work, together with their correlation coefficients, null-hypothesis probabilities referred to the goodness of the fit and statistical significance. The most relevant relations are discussed in Section \ref{sect:discussion}.}
\label{table:all_fitting_param}  
\centering 
\begin{tabular}{lclccccc}   
\hline\hline                
\multicolumn{3}{l}{Relation (y vs. x)}& Slope (A) &Intercept (B) &Pearson & 1-P$_{\rm null}$ & $\sigma$\\
\hline\hline    
$\Gamma$& vs.& $v_{\rm C\,\textsc{iv}}$       & $(-1.72 \pm 0.79)\times 10^{-4}$  & $1.93 \pm 0.19$  & -0.901 & 0.9992 & $3.2$\\
$\Gamma$& vs.&$\log(M_{\rm s, Edd})$& $(3.87  \pm 1.89)\times 10^{-1}$  & $0.96 \pm 0.65$  & 0.554  & 0.9944 &$2.5$\\
$\Gamma$& vs.&$\log(REW_{\rm C\,\textsc{iv}})$& $(-3.66 \pm 2.75)\times 10^{-1}$  & $2.79 \pm 0.40$  & -0.760 & 0.9825&$2.4$\\
$\Gamma$& vs.&$z$                             & $(4.62 \pm 2.58)\times 10^{-1}$   & $-0.71 \pm 1.67$ & 0.813  & 0.9794&$2.3$\\
$\Gamma$& vs.&$\log(L_{\rm x})$               & $(-1.10\pm4.07)\times 10^{-1}$    & $7.30 \pm 1.84$  & -0.147 & 0.7506&$1.1$\\
$\Gamma$& vs.&$\log(\lambda_{\rm Edd})$       & $(2.94 \pm 4.23)\times 10^{-1}$   & $2.31 \pm 0.10$  & 0.426  & 0.6287 &$<1$\\
$\Gamma$& vs.&$\log(M_{\rm BH})$              & $(4.57 \pm 4.13)\times 10^{-2}$   & $1.85\pm 3.88$   & 0.087  & 0.4305&$<1$\\
$\Gamma$& vs.&$\log(L_{\rm bol})$             & $(1.05 \pm 2.87)\times 10^{-2}$   & $-2.71 \pm 1.47$ & 0.172  & 0.0919&$<1$\\
\noalign{\smallskip}
\hline
\noalign{\smallskip}
$\log(L_{\rm x})$ & vs. & $z$                          & $(-2.48 \pm 1.52)\times 10^{-1}$  & $46.86 \pm 1.12$   & -0.403 & 0.8611& $1.5$\\
$\log(L_{\rm x})$ & vs. &$v_{\rm C\,\textsc{iv}}$      & $(6.90 \pm 4.53)\times 10^{-5}$   &  $45.36 \pm 0.12$  & 0.436  & 0.8534&$1.5$\\
$\log(L_{\rm x})$ & vs. &$\log(M_{\rm BH})$            & $(1.44\pm 2.78)\times 10^{-1}$    & $43.88\pm 2.60$    & 0.157  & 0.8345&$1.5$\\
$\log(L_{\rm x})$ & vs. &$\log(REW_{\rm C\,\textsc{iv}})$ & $(1.74 \pm 2.18)\times 10^{-1}$   & $44.98 \pm 0.31$   & 0.263  & 0.8144&$1.4$\\
$\log(L_{\rm x})$ & vs. &\logmseed & $(-1.74 \pm 1.31)\times 10^{-1}$  & $45.83 \pm 0.86$   & -0.350 & 0.7874&$1.3$\\
$\log(L_{\rm x})$ & vs. &$\log(L_{\rm bol})$             & $(413 \pm 2.60)\times 10^{-1}$    & $25.67 \pm 12.32$  & 0.395  & 0.7076&$1.1$\\
$\log(L_{\rm x})$ & vs. &$\log(\lambda_{\rm Edd})$       & $(1.50 \pm 2.93)\times 10^{-1}$   & $45.25 \pm 0.08$   & 0.142  & 0.0806&$<1$\\
\noalign{\smallskip}
\hline
\noalign{\smallskip}
$\log(REW_{\rm C\,\textsc{iv}})$ & vs. &$v_{\rm C\,\textsc{iv}}$  & $(1.58  \pm 0.47)\times 10^{-4}$  & $1.77\pm 0.13$  & 0.786  & 0.9999 & $3.7$\\
$v_{\rm C\,\textsc{iv}}$& vs. & \logmseed  & $(-1.64 \pm 6.43)\times10^{+2}$  & $(3.30 \pm 2.23)\times10^{+3}$    & -0.603 & 0.9934 &$2.5$\\
$\log(M_{\rm s, Edd})$& vs. & $z$                               & $(1.06  \pm0.15)$                & $-3.50 \pm 1.02$  & 0.867  & 0.9999 & $3.7$\\
$v_{\rm C\,\textsc{iv}}$& vs. & $z$                               & $(-2.03 \pm0.78)\times10^{+3}$   & $(1.09\pm0.51)\times10^{+4}$    & -0.613 & 0.9962 & $2.7$\\
\hline
\hline
\end{tabular}
\end{table*}

\section{Discussion}
\label{sect:discussion}
\subsection{Linking X-ray corona and accretion disc winds}
\label{subsect:gamma_vCIV}
 \begin{figure*}
\sidecaption
  \includegraphics[width=12cm]{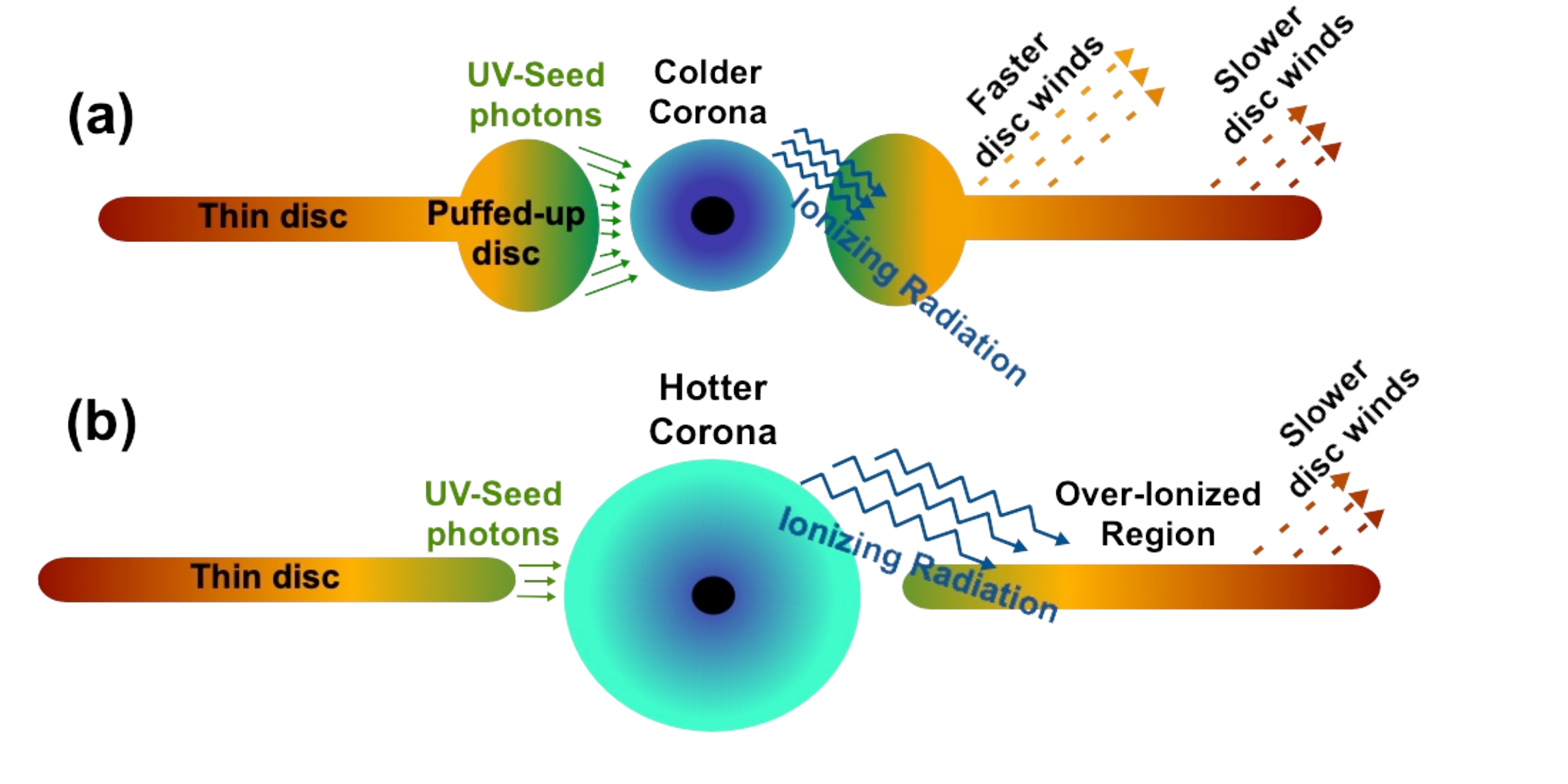}
     \caption{Scheme of the geometry of the AD--corona model proposed to provide a qualitative explanation for the observed $\Gamma - v_{\rm C\,\textsc{iv}}$ relation. Panel (a): AD--corona system in which the inner part of the AD is puffed up and the X-ray corona is cold, producing steep X-ray spectrum. In this configuration, faster AD winds can be launched from more internal AD regions. Panel (b): Standard AD and a hotter X-ray corona producing a canonical X-ray spectrum where only low-velocity winds can be launched from outer regions of the disc.}
     \label{fig:scheme}
\end{figure*}
We investigated the relation between the X-ray photon index and the velocities of the rest-frame UV disc winds in our sample of $L_{bol}>10^{47}$\,erg\,s$^{-1}$, $z>6$ QSOs and we discovered that they are related: the steeper the X-ray spectrum, the faster the $\rm C\,\textsc{iv}$ winds.

The interpretation of this measured relation is not straightforward but its presence may suggest a physical link between the disc--corona system and the terminal velocities of the AD winds. Current models of the AD wind and X-ray coronae for highly-accreting AGN \citep[e.g.][]{2018MNRAS.480.1247K,2018MNRAS.480.5184N,2023MNRAS.518.6065J}, assuming innermost hotter corona and outer AD configuration, can provide a qualitative explanation of how the $\Gamma$-$v_{\rm C\,\textsc{iv}}$ relation may originate. In this configuration, the physical properties and the relative geometry of the AD--corona system could play a crucial role in reproducing the $\Gamma - v_{\rm C\,\textsc{iv}}$ relation. In the case of a sufficiently high dimensionless mass accretion rate (ratio of the accretion rate, $\dot{M}$, to the Eddington accretion rate, $\dot{m}=\dot{M}/\dot{M}_{\rm Edd}$ where $\dot{M}_{\rm Edd}=L_{\rm Edd}/\eta c^2$), it has been suggested in different works \citep[e.g.][]{2015ApJ...805..122L,2016MNRAS.456.3929S,2019MNRAS.489..524K,2019ApJ...885..144J} that the inner part of a standard optically thick geometrically thin AD \citep{1973A&A....24..337S} puffs up (i.e. slim disc, \citealt{2019Univ....5..131C}), due to the high radiation-pressure of UV photons. The inner puffed-up part of the AD increases the EUV/soft X-ray flux seen by the corona (see Figure \ref{fig:scheme}a), leading to a much more efficient cooling of the corona and a steeper X-ray spectrum. Furthermore, a similar inner puffed-up component of the AD shields the standard part of the AD beyond it from the X-rays. This prevents significant over-ionization and allows the line-driven acceleration of $\rm C\,\textsc{iv}$-emitting gas from the more internal regions of the thin AD, leading to winds with higher velocities. Conversely, for lower accretion rates, the absent or reduced puffing of the inner part of the AD results in fewer UV-seed photons. Consequently, the corona cooling is lower and the outcoming X-ray continuum slope is flatter, but consistent with most AGN (see Figure \ref{fig:scheme}b). Accordingly, the harder ionizing X-ray radiation from the corona will not be shielded, over-ionizing the inner regions of the AD. Therefore, the winds will be launched at larger radii, with slower velocity.

According to this scenario, we would expect, together with the measured $\Gamma - v_{\rm C\,\textsc{iv}}$ relation, a $\Gamma - \lambda_{\rm Edd}$ relation; however, we do not recover it (see Section \ref{subsect:testing_accr_properties}). 

The properties of our sample resemble those of the narrow-line Seyfert\,1 galaxies (NLS1s; \citealp{1994ApJ...435L.125M,2002A&A...388..771C,Collin_2004}), although scaled up in mass and luminosity by at least two orders of magnitude. Actually, NLS1s host SMBHs accreting close to or above the Eddington limit \citep{Pounds1995,Komossa_2006,2016MNRAS.455..691J,10.1093/mnras/stx718,2015A&A...578A..28B}. They show very steep X-ray continuum slopes, possibly driven by enhanced EUV/soft X-ray emission \citep{1996A&A...305...53B,1997MNRAS.285L..25B,1999ApJS..125..297L,1999ApJS..125..317L} and high-velocity blueshifted C\,\textsc{iv} lines \citep{2000ARA&A..38..521S,2002ApJ...566L..71S,2000NewAR..44..511W,2004ApJ...611..107L}. The QSOs within our sample typically exhibit X-ray continuum slopes that are steeper compared to analogous QSOs at redshifts below 6 (but see e.g.  the case of PHL\,1092, \citealt{2009MNRAS.396L..85M,2012MNRAS.425.1718M}). Focusing on similar $M_{\rm BH}$ and $L_{\rm bol}$ ranges, the population of X-ray unabsorbed weak emission-line quasars (WLQ; e.g. \citealt[][]{Fan_1999,Diamond-Stanic_2009,Plotkin_2010}) show steep X-ray spectra \citep{Luo_2015} and velocity shifts of the $C\,\textsc{iv}$, $v_{\rm C\,\textsc{iv}} \sim 1000-8000$\,km/s \citep{Wu_2011,Wu_2012,Luo_2015}. Despite the similarities, to our knowledge, to date no evidence of a relation between $\Gamma$ and $v_{\rm C\,\textsc{iv}}$ has been reported for these categories of objects, nor for the population of AGN in general.

\begin{figure}
\centering
\includegraphics[width=0.9\columnwidth]{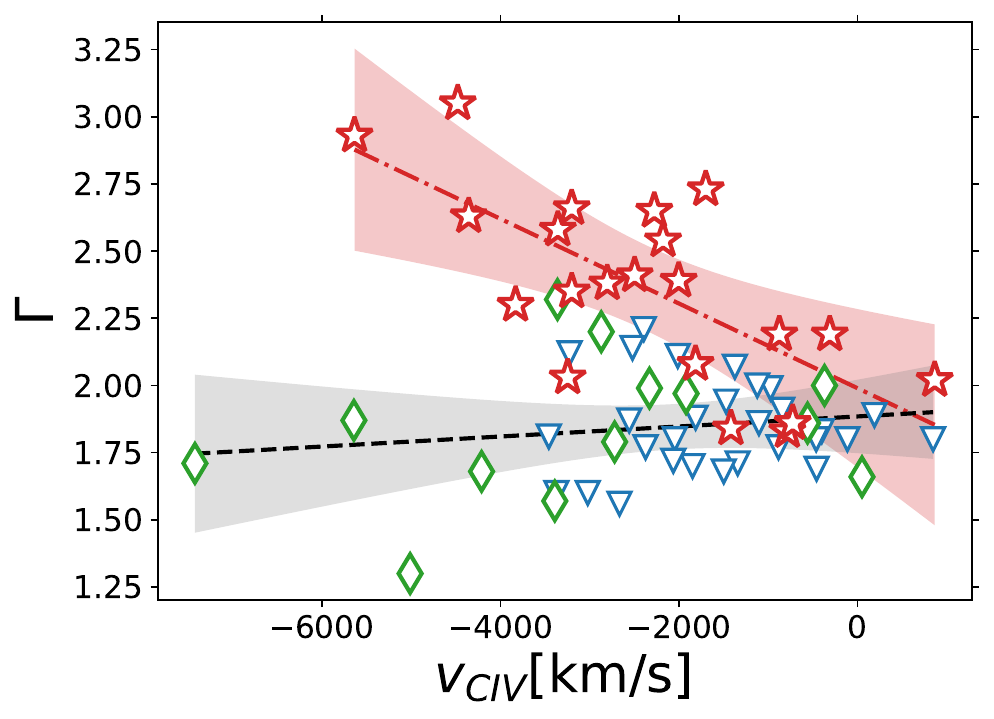}
\caption{$\Gamma$ vs. $v_{\rm C\,\textsc{iv}}$ for our sample of $z>6$ QSOs (red stars) compared with a sample of luminous QSOs at $z<6$: the WISSH QSOs at $z = 2 - 4$ \citep{Zappacosta2020} (green diamonds) and blue QSO at $z\simeq 3.0 - 3.3$ \citep{Nardini2019,Lusso2021} (blue inverted triangles). We report the linear relations we found for our sample of QSOs at $z>6$ (red dot-dashed line) and for the combined sample of lower $z$ QSOs (black dashed line). The shaded regions represent the combined 1$\sigma$ error on the slope and normalization.}
\label{fig:control_sample1}
\end{figure}

The relation between the X-ray continuum slope and the AD wind velocity has never been reported for AGN at $z<6$. In \citet{Zappacosta2020}, the $\Gamma$ vs. $v_{\rm C\,\textsc{iv}}$ relation was tested for a sample of $z\sim 2-4$ WISE/SDSS-selected Hyper-luminous QSOs (WISSH), but no correlation was recovered. In Figure \ref{fig:control_sample1} we report the $\Gamma$ vs. $v_{\rm C\,\textsc{iv}}$ relation for the \citet{Zappacosta2020} sample and the sample of $z\simeq 3.0 - 3.3$ QSOs selected from the Sloan Digital Sky Survey (SDSS) seventh Data Release (DR7) \citep{Nardini2019,Lusso2021}, along with our sample. We performed a correlation analysis over the combined sample of QSOs at $z<6$ using the same method described in Section \ref{subsect:fitting}, but we do not find a significant relation ($<1\sigma$ and Pearson correlation coefficient $\rho=0.162$). 
The absence of a similar relation for QSOs with similar $L_{\rm bol}$ ($\gtrsim10^{47}$\,erg\,s$^{-1}$) and $\lambda_{\rm Edd}$ at $z<6$, may be driven by a different $\dot{m}$. In particular, at a given $\lambda_{\rm Edd}$, the $z<6$ QSOs may have a more standard AD--corona configuration (see Figure \ref{fig:scheme}b) than $z>6$ QSOs, as a consequence of lower $\dot{m}$. This would imply higher radiative efficiencies and consequently higher BH spins at $z<6$. Thus, the measured $\Gamma$ vs. $v_{\rm C\,\textsc{iv}}$ relation may be a distinctive feature of $z>6$ QSOs, possibly indicating specific high $\dot{m}$ regimes ranging between the disc--corona configurations sketched in Figure \ref{fig:scheme}.

Within our sample (see Figure \ref{fig:main_results}), the HYPERION sources contribute with steeper X-ray spectra and faster and more powerful winds (i.e. with higher kinetic power, $\propto v_{\rm C\,\textsc{iv}}^3$; e.g. \citealt{2018A&A...617A..81V}), driving the observed relation.  This suggests that the HYPERION selection, based on \mseed\ and $L_{\rm bol}$, allows us to select the most extreme QSOs in terms of $\Gamma$ and AD winds (see Section \ref{subsect:SMBH_growth}). We calculated the average $v_{C\,\textsc{iv}}$ for the \citealt{Yang2021} ($z>6.3$) and \citealt{Farina2022} ($z>6.0$) QSOs samples (avoiding duplication with the HYPERION sources), finding $\sim -1540$ km/s and $\sim -1700$ km/s, respectively. Matching the HYPERION sample to the same redshift ranges of \citet{Yang2021} and \citet{Farina2022}, we found an average $v_{C\,\textsc{iv}}\sim -3100$\, km/s and $\sim -2800$\,km/s, respectively. 
Interestingly, the low $\Gamma=1.83$, resulting from the stacking spectral analysis of \citet{2017A&A...603A.128N} for ten QSOs with $5.7<z<6.1$, may be explained according to our reported relation if the sources employed in their stacking analysis had, on average, a very low $v_{\rm C\,\textsc{iv}}$ (from our relation we would expect $\sim600$\,km/s, see Figure \ref{fig:main_results}).

\subsection{Testing the presence of an accretion rate and X-ray radiative power dependence}
\label{subsect:testing_accr_properties}
We analysed the possible presence of a relation between $\lambda_{\rm Edd}$ and $\Gamma$ in the QSOs in our sample. 
The analysed sample of QSOs covers a limited range size for $\lambda_{\rm Edd}$, with a limited number of $z>6$ sources showing a large dispersion in terms of $\Gamma$. It should be considered that there are large uncertainties on the measure of $\lambda_{\rm Edd}$ (related to the measure of $M_{\rm BH}$). Moreover, while in sub-Eddington regimes $L_{\rm bol} \propto \dot{m}$, at higher accretion rates this is not true anymore, since their relation is strongly affected by the radiative efficiency of the accretion, also related to the spin of the SMBH \citep{2019Univ....5..131C,2011arXiv1108.0396S,2014ApJ...784L..38M}. In particular, in the slim disc (puffed-up disc) scenario, $\dot{m}$ saturates due to photon trapping and advection, which cause a decrease in the radiative efficiency. Because the radiative efficiency is dependent on the spin of the SMBH, for higher SMBH spin, $\dot{m}$ saturates more \citep[see Fig.\,1 of ][]{2014ApJ...784L..38M}. In our sample we analysed a very narrow range of $\lambda_{\rm Edd}$ which is also in the region of the transition between the standard geometrically thin disc and the geometrically thick disc (slim disc). In this situation $\lambda_{\rm Edd}$ may not be a good proxy for $\dot{m}$ and this could be another reason why we do not see a $\Gamma - \lambda_{\rm Edd}$ relation. All these prevented us from obtaining firm constraints on the presence of a possible relation between the two quantities. However, we can compare this result and the distribution of HYPERION sources in this parameter space with that reported for samples of lower-$z$ objects reported in the literature.

\begin{figure}
\centering
\includegraphics[width=0.9\columnwidth]{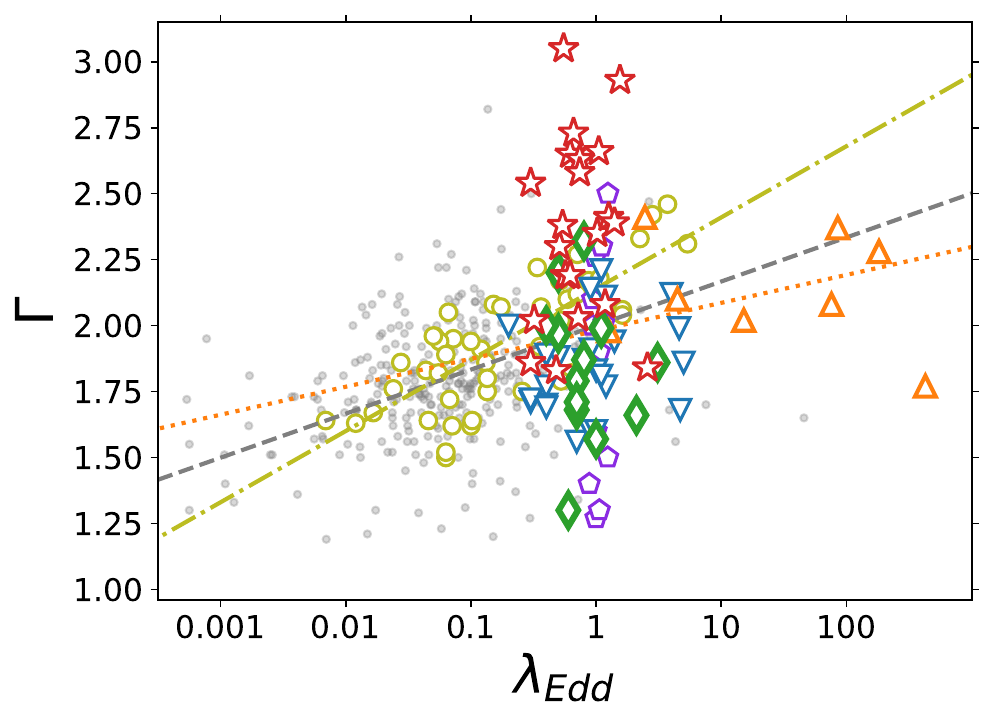}
\caption{$\Gamma$ vs. $\lambda_{\rm Edd}$ for our sample of $z>6$ QSOs (red stars) compared with the WISSH sample of QSOs at $z = 2 - 4$ (green diamonds, \citealt{Zappacosta2020}), the SDSS DR7 blue QSOs at $z\simeq 3.0 - 3.3$ (blue inverted triangles, \citealt{2023A&A...677A.111T}), local hyper-Eddington AGN (orange triangles, \citealt{2023MNRAS.519.6267T}), radio-quiet super-Eddington accreting QSOs at $0.4 \leq z \leq 0.75$ (purple, pentagons \citealt{2022A&A...657A..57L}), a sample of $z<0.4$ AGN (yellow circles, \citealt{2021ApJ...910..103L}) and the local type\,1 AGN belonging to the BASS sample (grey points, \citealt{2017MNRAS.470..800T,2017ApJS..233...17R}).  We also report the linear relations from \citealt{2017MNRAS.470..800T} (grey dashed line), \citealt{2021ApJ...910..103L} (yellow dot-dashed line) and \citealt{2023MNRAS.519.6267T} (orange dotted line).}
\label{fig:control_sample2}
\end{figure}
We compared the $\Gamma$ vs. $\lambda_{\rm Edd}$ relation found in our sample with other samples of $L_{\rm bol}>10^{47}$\,erg\,s$^{-1}$) QSOs at lower $z$ extending over wider ranges of $\lambda_{\rm Edd}$. We included the WISSH sample of QSOs at $z = 2 - 4$ from \citet{Zappacosta2020}, the blue QSOs at $z\simeq 3.0 - 3.3$ selected from the SDSS DR7 from \citet{2023A&A...677A.111T}, the local hyper-Eddington accreting AGN \citep[$1.2 \leq \lambda_{\rm Edd} \leq 468$,][]{2023MNRAS.519.6267T}, the radio-quiet AGN at $0.4 \leq z \leq 0.75$ accreting at $0.9\leq \lambda_{\rm Edd}\leq1.1$ \citep{2022A&A...657A..57L}, a sample of $z<0.4$ AGN with reverberation mapping measurements which includes super-Eddington accreting AGN and sub-Eddington accreting AGN \citep{2021ApJ...910..103L} and the sample of unobscured ($N_H< 10^{22} \rm cm^{-2}$), radio quiet, type\,1 AGN belonging to the Swift/BAT AGN Spectroscopic Survey (BASS) from \citet{2017MNRAS.470..800T} with a median redshift of $0.035$. Remarkably, $\sim 80\%$ of the QSOs in our sample are located above the $\lambda_{\rm Edd}$ vs. $\Gamma$ relations (see upper left panel of Figure \ref{fig:other_relation} and Figure \ref{fig:control_sample2}) presented in the literature for local AGN \citep[e.g.][]{2017MNRAS.470..800T,2021ApJ...910..103L,2023MNRAS.519.6267T}. 

We note that the values of $\Gamma$ in the different samples wew obtained sampling different rest-frame energy ranges. However, the $\Gamma$ values from the BASS sample \citep{2017ApJS..233...17R} and for the sample from \citet{2023MNRAS.519.6267T} were obtained at $\gg 10$\,keV rest-frame energies, similarly to the HYPERION sample. Furthermore, in \citet{Zappacosta2023}, fits performed from $2$\,keV (i.e. rest-frame low-energy bound similar to that of the HYPERION QSOs), are reported for the WISSH QSOs analysed in \citet{Zappacosta2020}. The reported $\Gamma$ values are comparable to those obtained also accounting for softer X-ray photons ($<2$\,keV). This makes the comparison between the $\Gamma$ measurements in different samples reliable, and strengthens the evidence for the steepening of the X-ray spectra of the first QSOs reported by \citet{Zappacosta2023}.

We do not find a relation between $L_{\rm x}$ and $v_{\rm C\,\textsc{iv}}$ for the QSOs at $z>6$ analysed in this work. We investigated this relation at a fixed UV luminosity (i.e. in a range smaller than $\sim0.5$\,dex) as in \citet{Zappacosta2020}. Thus, we excluded SDSS\,J0100+2802 because it is an outlier in terms of $L_{\rm bol}$ (see Table \ref{table:sources}). At a fixed UV luminosity, luminous QSOs at Cosmic Noon show a $L_{\rm x}$ vs. $v_{\rm C\,\textsc{iv}}$ anti-correlation \citep{Zappacosta2020}. However, within the uncertainties, we are in agreement with the relation found by \citet{Zappacosta2020} (see upper right panel of Figure \ref{fig:other_relation}). We need to enlarge the dataset, to broaden the statistics and the interval of probed $v_{\rm C\,\textsc{iv}}$ and to test the presence of a $L_{\rm x}$ vs $v_{\rm C\,\textsc{iv}}$ relation for $z>6$ QSOs on a more solid basis.
\subsection{Linking the X-ray corona with the SMBH mass growth history}
\label{subsect:SMBH_growth}
We tested the $\Gamma$ vs. \logmseed and $\Gamma$ vs. $z$ relations, confirming, on a larger sample, the earlier claims by \citet{Zappacosta2023} of a $\Gamma$ - \mseed\ dependence. In our analysis we found these relations to be moderate, with $<3\sigma$ significance, with the $\Gamma$ - \mseed\ dependence slightly more significant than that of the $\Gamma - z$. To check whether the $\Gamma - z$ is affected by the potential selection bias of $z>7$ QSOs we fitted the $6<z<7$ QSOs recovering a similar relation to what obtained when fitting the whole sample, with a slightly lower statistical significance ($2.1\sigma$). This suggests that the result is not influenced by the smaller $z>7$ sample.

As argued in \citealt{Zappacosta2023}, the possibility that these sources experienced a redshift-dependent steepening of the spectra seems unlikely as it must happen in $0.2-0.3$\,Gyr (see lower right panel of Figure \ref{fig:other_relation}). This timescale is too short for any cosmological evolution to take place. The hypothesis that the steepening of $\Gamma$ is dependent on \mseed\ (i.e. driven by the fast SMBH growth experienced by these QSOs during their formation history) seems to be more likely.

However, $\Gamma$ is an instantaneous quantity; it can quickly change in response to a change in the accretion state phase \citep[e.g.][]{2009MNRAS.399.1597S,2017MNRAS.470..800T,Serafinelli2017}. The existence of a tentative relation between \mseed, a proxy of the integrated growth history of the SMBH, is therefore intriguing. Between the two most popular pathways to grow $z>6$ SMBHs, we consider the super Eddington growth as a more likely explanation of the relation with $\Gamma$ as it is related to the accretion mechanism rather than to the initial seed BH mass. This would imply that these sources have been typically accreting at super-Eddington rates over their lifetime. This is at variance with their $\lambda_{\rm Edd}$ values, which are on average $< 1$. Hence, it is possible that these sources have experienced a hybrid scenario whereby their growth is favoured by initial massive seeds and by a fast SMBH growth that is representative of the currently observed $\lambda_{\rm Edd}$.

The $\Gamma$ vs. \logmseed\ relation may not be a fundamental relation; rather, it is being derived from a third parameter that is by itself truly linked to \mseed. Furthermore, the existence of $\Gamma$ vs. $v_{\rm C\,\textsc{iv}}$ and $\Gamma$ vs. \logmseed\ relations implies that $v_{\rm C\,\textsc{iv}}$ may depend on \mseed. We estimated a moderate significance ($2.5\sigma$, see Table\,\ref{table:all_fitting_param}) of the correlation between $v_{\rm C\,\textsc{iv}}$ and \mseed\ in our sample. However the significance of this relation (which does not depend on the X-ray parameters) can be verified on a much larger sample of QSOs. 

\begin{figure*}
   \centering
   \includegraphics[width=0.9\columnwidth]{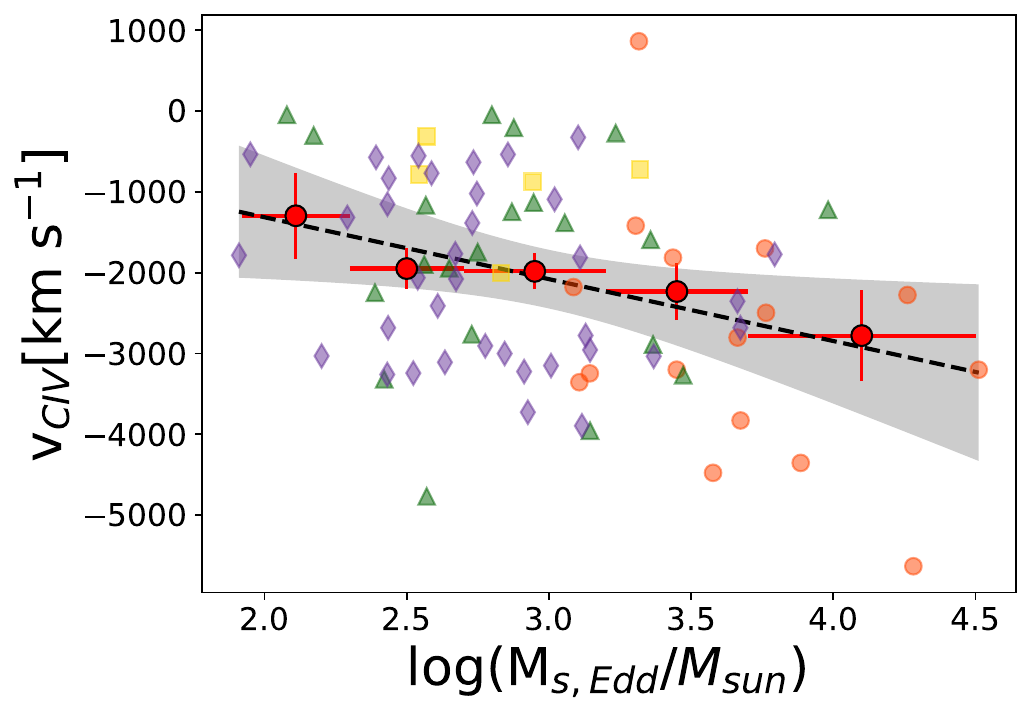}
   \includegraphics[width=0.92\columnwidth]{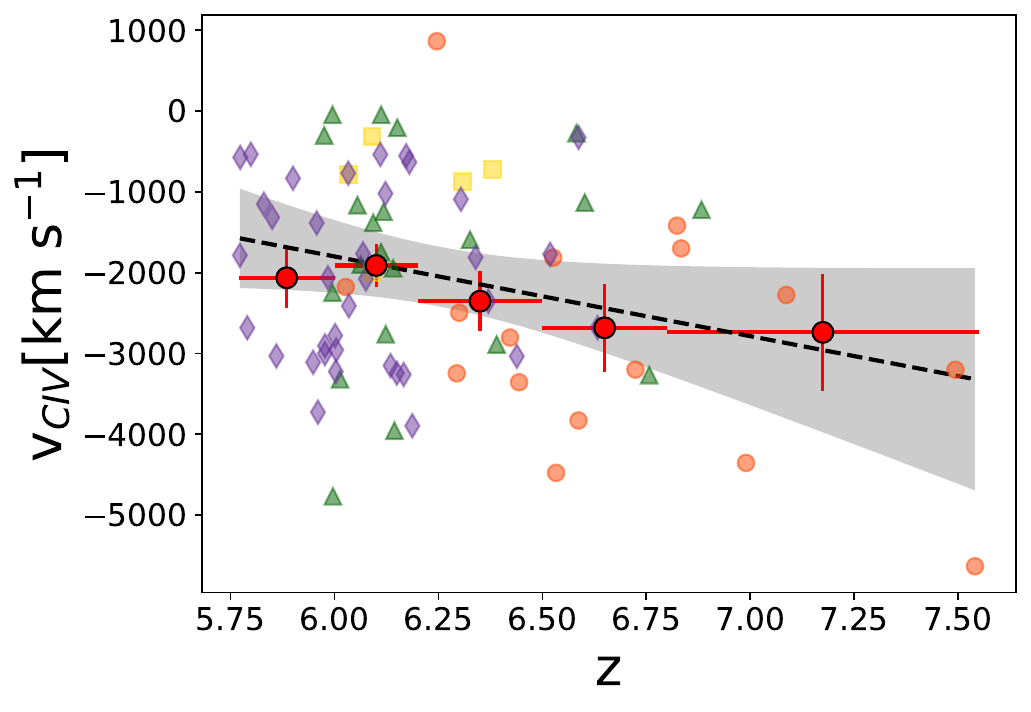}
   \caption{Plot of the relations which do not involve X-ray data extending the data sample. Left panel: $v_{\rm C\,\textsc{iv}}$ vs. $\log(M_{\rm s, Edd})$ linear relation; Right panel: $v_{\rm C\,\textsc{iv}}$ vs. $z$ linear relation. The orange circles and the yellow squares represent the HYPERION and non-HYPERION QSOs in our sample, respectively. The green triangles and the purple diamonds represent the QSOs from \citealt{Farina2022} and \citealt{Mazzucchelli2023}, respectively. the red circles with errorbars correspond to the median value for each bin. The dashed black lines are the linear regressions while the shaded regions represent the combined 1$\sigma$ error on the slope and normalization (see Table \ref{table:fitting_param_ext}).}
   \label{fig:relation_ext}
\end{figure*}

We also tested the relations between $z$, \mseed, and $v_{\rm C\,\textsc{iv}}$, which do not involve X-ray data, on an extended sample of $\sim80$ QSOs, which includes, along with the sources of this work, the sources with the most recent reliable $v_{\rm C\,\textsc{iv}}$ measurements derived by \citet{Farina2022} and \citet{Mazzucchelli2023} (see Figure \ref{fig:relation_ext}). The sample from \citet{Farina2022} includes 31 bright QSOs at $5.78 < z < 7.54$, while the sample from \citet{Mazzucchelli2023} consists of 37 luminous QSOs at $z \geq 6$ with high signal-to-noise ratio VLT/X-Shooter spectra (avoiding duplication with the HYPERION sample), acquired in the enlarged ESO Large Programme XQR-30 \citep[P.I. D'Odorico,][]{2023MNRAS.523.1399D}. For the comparison with our sample, we did not considered duplication in these samples. We applied the same fitting method described in Section \ref{sect:sample_and_data_red} to the data. The best--fitting parameters for the $v_{\rm C\,\textsc{iv}}$vs. $z$ and $v_{\rm C\,\textsc{iv}}$ vs. \mseed\ relations are reported in Table \ref{table:fitting_param_ext}. Figure \ref{fig:relation_ext} shows the plots of the relations we analysed in the extended sample. We also reported the median values computed in bins of \mseed\ and $z$ (red stars with errorbars in Figure \ref{fig:relation_ext}). We chose to have the bins containing at least five objects. The binned data exhibit the same trend of the fits. We found that, expanding the sample, the strength of the $v_{\rm C\,\textsc{iv}}$ vs. \mseed\ relation increases and the statistical significance increases to $3\sigma$ (see left panel of Figure \ref{fig:relation_ext} and Table \ref{table:fitting_param_ext}). We checked whether the $v_{\rm C\,\textsc{iv}}$ vs. \mseed\ and the $v_{\rm C\,\textsc{iv}}$ vs. $z$ relations are driven by the $z>7$ QSOs subsample by excluding them from the fit of the extended sample (including the QSOs from \citealt{Farina2022} and \citealt{Mazzucchelli2023}). We found relations that are completely in agreement within the errors with those obtained considering the total sample.
\begin{table*}
\caption{Best-fit relations for the extended sample analysed here with their correlation coefficients, null-hypothesis probabilities referred to the goodness of the fit and statistical significance.}
\label{table:fitting_param_ext}  
\centering 
\begin{tabular}{lclccccc}   
\hline\hline                
\multicolumn{3}{l}{Relation (y vs. x)}& Slope (A) &Intercept (B) &Pearson & 1-P$_{\rm null}$ & $\sigma$\\
\hline\hline          
$v_{\rm C\,\textsc{iv}}$ & vs. & $\log(M_{\rm s, Edd})$ & $(-7.94\pm 2.45)\times10^{+2}$ & $416 \pm 73$ & -0.382 & 0.9988 &$3.04$\\
$v_{\rm C\,\textsc{iv}}$& vs. & $z$  & $(-9.81\pm 4.24)\times10^{+2}$ & $4193 \pm 2654$ & -0.286 & 0.9868&$2.22$\\
\hline\hline
\end{tabular}
\end{table*}
Past works analysing the average $v_{\rm C\,\textsc{iv}}$ of $z>6$ QSOs, claimed a possible correlation between the $v_{\rm C\,\textsc{iv}}$ and $z$ \citep{Schindler2020,Yang2021}. However, by considering the extended sample, which has a factor of 2.4 more sources than these samples, we found a marginal ($2.2\sigma$ significance) correlation between $v_{\rm C\,\textsc{iv}}$ and $z$ (see right panel of Figure \ref{fig:relation_ext}, Table \ref{table:fitting_param_ext}). 

The existence of a more significant \mseed\ dependence and the insufficient timescale argument could suggest that the real relation could be between $v_{\rm C\,\textsc{iv}}$ and \mseed. Furthermore, as postulated for the $\Gamma$ vs. \logmseed\ case, it is possible that the growth pathway for these sources includes a massive seed followed by high (but not super-Eddington) accretion rates (i.e. $\lambda_{\rm Edd}>0.5$) for a significant time of the evolution during which powerful winds may be launched \citep{2018MNRAS.480.1247K,Jiang2019,Okuda_2021,2021A&A...646A.111L}.

Finally, we note that the definition of \mseed\ (see Equation \ref{eq:mseed}) involves a dependence on several parameters, including the details of the accretion flow; of the link between the AD--corona system; and of the link between the spin, the efficiency, and the possible impact from AD-driven outflows. A detailed examination of the interconnection among the various parameters involved within \mseed, along with a theoretical effort, is essential, but this is beyond the aim of this work. 
\section{Conclusions}
\label{sect:conclusion}
We investigated the relations between the coronal X-ray properties, the velocities of the AD winds and the properties regarding the physics and growth of the AD and SMBH of a sample of luminous ($L_{bol}>10^{47}$\,erg\,s$^{-1}$) QSOs at $z>6$. The sample is composed of 16 sources belonging to the HYPERION sample \citep{Zappacosta2023} and five QSOs from \citealt{Connor2019}, \citealt{Vito2019}, and \citealt{Pons2020} for which high--quality X-ray observations and $v_{\rm C\,\textsc{iv}}$ measurements are available.

Our main finding can be summarized as follows:
\begin{itemize}
    \item We find a significant ($>3\sigma$) anticorrelation between $\Gamma$ and $v_{\rm C\,\textsc{iv}}$ (see Figure \ref{fig:main_results}). This relation indicates a connection between the AD--corona configuration and the kinematics of AD winds. We explain it qualitatively in inner corona--outer AD models as a consequence of the transition between slim and standard discs at high accretion regimes. The formation of a inner geometrically puffed up AD (typical of high accretion flows) plays a role in producing steeper X-rays (as a consequence of coronal cooling induced by the high UV flux from this region) and in launching fast ionized AD winds from inner regions (which are shielded and hence not over-ionized by the central X-ray source flux).
    \item The QSOs in our sample and especially the HYPERION QSOs, generally show steeper X-ray continuum slopes compared to the canonical values, confirming the early claims \citep{Vito2019,Wang2021a,Zappacosta2023} according to which the first QSOs have intrinsically different X-ray properties. Roughly $80\%$ of the QSOs of the analysed sample is located above the $\lambda_{\rm Edd}$ vs. $\Gamma$ relations (see upper left panel of Figure \ref{fig:other_relation} and right panel of Figure \ref{fig:control_sample2}) presented in the literature for local AGN \citep[e.g.][]{2017MNRAS.470..800T,2021ApJ...910..103L,2023MNRAS.519.6267T}. This is a further evidence of the steepening of the X-ray spectra of $z>6$ QSOs. However, our data does not allow us to discriminate between a simple power law or a cut-off power law model. Because $\Gamma$ and low $E_{\rm cut}$ are both tracers of relatively cold coronae, our result could suggest low coronal temperatures for the QSOs in our sample.
    \item We do not recover a statistically significant relation between $\lambda_{\rm Edd}$ and $\Gamma$ in the QSOs in our sample (see upper left panel of Figure \ref{fig:other_relation}). However, we note that the analysed sample of QSOs allowed us to investigate a restricted range for $\lambda_{\rm Edd}$, with a limited number of $z>6$ sources. Moreover, the $\lambda_{\rm Edd}$ parameters for the QSOs in our sample is located in the transition region between the standard geometrically thin AD and the geometrically thick AD. In this situation $\lambda_{\rm Edd}$ may not be a reliable proxy for the real accretion rate. All these prevented us from drawing firm conclusions on the existence of a relation between  $\lambda_{\rm Edd}$ and $\Gamma$ in the QSOs in our sample.
    \item We tested the link between $L_{\rm x}$ and $v_{\rm C\,\textsc{iv}}$, finding a mild relation ($1.5\sigma$, see upper right panel of Figure \ref{fig:other_relation}). Within the uncertainties, there is a marginal agreement with the trend of the relation found by \citet{Zappacosta2020}, for a sample of analogous QSOs at Cosmic Noon. We need an expansion of the dataset, enhancing the statistics, to establish the existence of a $L_{\rm x}$ vs. $v_{\rm C\textsc{iv}}$ relation.
    \item A moderately significant dependence is found between $\Gamma$ and $z$, and between $\Gamma$ and \mseed, with a statistical significance of $2.3\sigma$ and $2.5\sigma$, respectively (see lower panels of Figure \ref{fig:other_relation}). Given that a redshift evolution of $\Gamma$ is highly unlikely to happen in the $0.2-0.3$\,Gyr corresponding to the redshift range in our sample (see lower right panel of Figure \ref{fig:other_relation}), a dependence with \mseed\ is more likely. This could imply that these sources can preferentially acquire their large SMBH masses via accretion at high rates.
    \item Similarly to the previous relation, we report a significant anticorrelation also between $v_{\rm C\,\textsc{iv}}$ and \mseed\ on a much larger sample regardless of their X-ray properties (see Figure \ref{fig:relation_ext}). 
    \item We report a significant ($>3\sigma$) $REW_{\rm C\,\textsc{iv}}$ vs. $v_{\rm C\,\textsc{iv}}$ anticorrelation in our sample of $z>6$ (see Figure \ref{fig:main_results2}), confirming the well--established trend in lower-$z$ QSOs \citep{2011AJ....141..167R,2014Natur.513..210S,2017MNRAS.464.3431H,2018A&A...617A..81V,2020A&A...644A.175V,Schindler2020,matthews2023disc}.
\end{itemize} 
A future extensive examination of the relations presented in this work is fundamental, from the theoretical and the observational points of view. On one hand, a thorough theoretical analysis of the intricate connections among the different parameters involved in \mseed\ it is necessary to establish the existence of a link between \mseed\ and the SMBH mass growth history of high-$z$ QSOs. On the other hand, it is crucial to expand the dataset by incorporating additional sources, especially those with a wider range of \mseed\ values. To this end, we will take advantage of our recently accepted \xmm Large programme observations (604\,ks, P.I. L. Zappacosta) of more hyper-luminous QSOs at $z>6$ significantly extending \mseed\ at the lowest values, by more than doubling the number of sources with \mseed\ $<1000M_{\odot}$. Moreover, we will jointly exploit state-of-the-art instrument (e.g. {\it Chandra}, JWST), possibly expanding the targets at even higher $z$ \citep[e.g. thanks to {\it Euclid},][]{2024NatAs...8..126B}.

The next-generation X-ray observatory such as the Advanced X-Ray Imaging Satellite \citep[{\it AXIS},][]{2023arXiv231100780R} will be pivotal to determine the nature of the seeds of these earliest growing SMBH at $z>6$ \citep[e.g.][]{2024Univ...10..276C}. Moreover, the Wide Field Imager \citep[WFI,][]{2013arXiv1308.6785R} large collecting area of the Advanced Telescope for High-Energy Astrophysics ({\it Athena}) mission, will play a crucial role by significantly minimizing the observing times. On the one hand this will increase the X-ray spectral quality and on the other hand it will permit the observation of more sources allowing a statistically more reliable characterization of the population of $z>6$ QSOs.

\begin{acknowledgements}
      This work is based on observations obtained with the ESA science mission {\it XMM-Newton}, with instruments and contributions directly funded by ESA Member States and the USA (NASA). We thank the \xmm Science Operation Centre for the prompt support and advise for the scheduling and optimization of the {\it XMM}-HYPERION program. The authors acknowledge financial support from the Bando Ricerca Fondamentale INAF 2022 Large Grant “Toward an holistic view of the Titans: multi-band observations of $z>6$ QSOs powered by greedy supermassive black holes" and the anonymous referee for the useful suggestions which helped in improving the manuscript. LZ, EP, FN, AL acknowledge support from the HORIZON-2020 grant “Integrated Activities for the High Energy Astrophysics Domain" (AHEAD-2020), G.A. 871158. GM thanks grant PID2020-115325GB-C31 funded by MCIN/AEI/ 10.13039/501100011033 for support. FT and EP acknowledge funding from the European Union - Next Generation EU, PRIN/MUR 2022 2022K9N5B4. SC acknowledges support from the European Union (ERC, WINGS, 101040227). POP acknowledges financial support from the french national space agency (CNES) and the High Energy programme from the french National Centre for Scientific Research (PNHE/CNRS). AB, MB, SC, VD, FF, CF, SG, VT, MN, LZ acknowledge support from the European Union - Next Generation EU, PRIN/MUR  2022 2022TKPB2P - BIG-z. MB acknowledges support from INAF project 1.05.12.04.01 - MINI-GRANTS di RSN1 "Mini-feedback" and from UniTs under FVG LR 2/2011 project D55-microgrants23 "Hyper-gal". A very special thanks from AT to the Hematology Department in the "Policlinico Umberto I" and to the Hematology Department and the Hematology Day Hospital in the "Policlinico Universitario Campus Bio-Medico". They saved my life and without them I would not have been able to complete this work.
\end{acknowledgements}

\bibliographystyle{aa} 
\bibliography{biblio} 
\begin{appendix} 

\end{appendix}
\end{document}